\documentclass[12pt]{article} 
\usepackage{graphicx,floatflt,amssymb,rotate}
\usepackage{mathrsfs}
\usepackage{amssymb}
\usepackage{amsmath}
\usepackage{color}
\usepackage{graphicx}
\usepackage{subfigure}
\usepackage{multirow}
\usepackage{float}
\usepackage{pstricks}
\usepackage[toc,page]{appendix}
\usepackage{pifont}
\usepackage{braket}
\usepackage{bbold}

\newcommand{\cmark}{\text{\ding{51}}}
\newcommand{\xmark}{\text{\ding{55}}}
\newcommand{\beq}{\begin{equation}}
\newcommand{\eeq}{\end{equation}}
\newcommand{\bea}{\begin{eqnarray}}
\newcommand{\eea}{\end{eqnarray}}

\def\nn{\nonumber}

\def\ra{\rightarrow}

\def\eg{ {\em e.g.,\ }}

\begin{document}
\vskip 30pt  
 
\begin{center}  
{\Large \bf Exploring the extended scalar sector with resonances in vector boson scattering
} \\
\vspace*{1cm}  
\renewcommand{\thefootnote}{\fnsymbol{footnote}}  
{{\sf Najimuddin Khan${}^{1}$\footnote{email: phd11125102@iiti.ac.in}},
{\sf Biswarup Mukhopadhyaya${}^{2}$\footnote{email: biswarup@hri.res.in}},
{\sf Subhendu Rakshit${}^{1}$\footnote{email: rakshit@iiti.ac.in} and }
{\sf Avirup Shaw${}^{2}$\footnote{email: avirup.cu@gmail.com}} 
} \\  
\vspace{10pt}  
{${}^{1}${\em Discipline of Physics, Indian Institute of Technology Indore,\\
 Khandwa Road, Simrol, Indore - 453~552, India}\\
${}^{2}${\em Regional Centre for Accelerator-based Particle Physics, Harish-Chandra Research Institute,
\\ HBNI, Chhatnag Road, Jhusi, Allahabad - 211~019, India}}
\normalsize  
\end{center} 

\begin{abstract}
We show that the study of scalar resonances at various vector boson
scattering processes at the Large Hadron Collider can serve as
a useful tool to distinguish between different extensions
of the scalar sector of the Standard Model. The recent measurement of
the Higgs boson properties leaves enough room for the extended scalar
sectors to be relevant for such studies. The shape of the resonances,
being model dependent, can shed light on the viable parameter space of
a number of theoretical models.
\end{abstract}
\vskip 5pt \noindent  
\texttt{PACS Nos.:~11.55.-m, 12.60.-i, 14.80.Cp} \\  
\texttt{Key Words:~Unitarity, Vector boson scattering, Resonance production}   
\setcounter{footnote}{0}  
\renewcommand{\thefootnote}{\arabic{footnote}}  

\section{Introduction}
Identifying the sector responsible for electroweak symmetry
breaking~(EWSB) has all along been the pre-condition for putting the
final stamp of approval on the Standard Model~(SM). The
discovery~\cite{Aad:2012tfa, Chatrchyan:2012xdj} of a largely SM-like
Higgs boson at the Large Hadron Collider (LHC) has, for all practical
purpose, accomplished this task. The measurement of properties of
this scalar boson is consistent with the minimal choice of the scalar
sector, namely, a single complex doublet. However, the data still
allow an extended scalar sector, which, in turn, can accommodate a
more elaborate mechanism for EWSB. One immediate extension of this
kind is the presence of either additional doublet(s) or scalars
belonging to some higher representation of SU(2). Even a marginal role
of such scalars can in principle be probed in the next round(s) of
experiments, utilising their interaction with the electroweak gauge
bosons.\footnote {It should be noted that the new scalars may not
  always participate in EWSB, \eg as in the inert doublet model
  \cite{Deshpande:1977rw}.}

If indeed there are additional scalars coupling to the W-and Z-bosons,
longitudinal vector boson scattering~(VBS) including scalar exchanges
should provide a complementary way to direct search methods to probe
the scalar sector. In the SM, the Higgs boson helps preserve the
unitarity of the $S$-matrix for the longitudinal electroweak vector
boson scattering $V_L V_L \ra V_L V_L$ (where $V\equiv
W^{\pm}~{\rm or}~Z$). The Higgs boson mediated
diagram precisely cancels the residual $s$-dependence (where $s$ is
the square of the energy in the centre-of-mass frame), thus taming the
high energy behaviour of the cross-section
appropriately~\cite{Dawson:1998yi}. With an extended scalar sector,
the preservation of unitarity could be a more complex process. Several
factors then modify the $\sqrt{s}$-dependence of the $V_{L}-V_{L}$
scattering.  The first of these is the modification, albeit small, of
the strength of the 125~GeV scalar to gauge boson pairs. Secondly, the
extent of the influence of other scalars present in an extended
scenario depends on their gauge quantum numbers and on the theoretical
scenario in general.  Thirdly, the observed mass of the 125~GeV scalar
makes it kinematically impossible for it to participate as an
s-channel resonance in $V_{L}-V_{L}$ scattering processes.  However,
such resonant peaks may in general occur when heavier additional
scalars enter into the arena.

One can thus expect that the $\sqrt{s}$-dependence of $V_{L}-V_{L}$
scattering cross-sections will be modified with respect to
SM-expectations as a result of the above effects. Such modifications
have been formulated in terms of certain general parameters in some
recent studies~\cite{Bhattacharyya:2012tj, Choudhury:2012tk,
  Chang:2013aya, Cheung:2008zh}. An apparent non-unitarity of the
scattering matrix may be noticed here when the SM-like scalar with
modified interaction strength is participating as the only
scalar. However, unitarity is restored once the full particle spectrum
is taken into consideration.  We emphasize here that the three effects
mentioned above leave the signature of the specific non-standard EWSB
sector in the modified energy-dependence, so long as the new scalars
have their masses within or about the TeV-range.

The aim of the present work can thus be summarised as follows.  Making
use of the resonant peaks in various $V_{L}-V_{L}$  scattering processes, we illustrate that it may
be possible to distinguish between different extensions of the scalar
sector, once the high-energy run of the LHC continues long enough. The
shapes of the energy-dependence curves, especially the presence of
resonant peaks, can shed light on the relevant scalar spectra of these
models.  We use for illustration some popular extensions like the
Type-II two Higgs doublet model (2HDM) and real as well as complex
Higgs triplet models (HTM). It is shown in the ensuing study how the
$\sqrt{s}$-dependence of the cross-sections reflect the
characteristics of each of these scenarios so long as the additional
scalars lie within about 2 TeV. This supplements rather faithfully
other LHC-based phenomenology, and thus spurs the improvement of
techniques to extract the dependence of $V_{L}-V_{L}$ scattering
cross-sections on $\sqrt{s}$.  We also present analytical
expressions for $V_{L}-V_{L}$ scattering amplitudes in these otherwise
well-motivated models. To the best of our knowledge, these full
expressions have not yet been presented in the literature.

In ref.\;\cite{Cheung:2008zh} the authors have studied this kind of
exercise by parametrising the coupling between SM Higgs and pair of
weak gauge bosons without introducing any new scalar, as a result the
scattering amplitude grow after the light Higgs pole due to incomplete
cancellation of the bad high-energy behaviour terms. On the other
hand, we have studied this issue in different models from a new
perspective, keeping in mind that unitarity is respected by the models
considered in our analysis. Consequently, the
$\sqrt{s}$-dependence of the cross-sections does not display any
ungainly growth; however, the specific character of the augmented
scalar sector is captured, basically through the occurrence of
resonant peaks, and from the invariant mass distributions in the
neighbourhood of the peaks.

In section~\ref{EssenPrinPl}, we outline the basic principles involved
in $V_{L}-V_{L}$ scattering.  The extended scalar sectors specifically
covered in this study have been summarised in
sections~\ref{NewModel}. In section~\ref{Nemeresult}, we present our
analysis and numerical results. We conclude in
section~\ref{conclusionSUM}.

\section{$V_L -V_L$ scattering: the essential points}\label{EssenPrinPl}
The differential cross-section for a $2 \rightarrow 2$ scattering
process is \beq\label{dexse}
\frac{d\sigma}{d\Omega}=\frac{1}{64\pi^2s}\frac{|p_f|}{|p_i|}|{\cal
  M}|^2, \eeq where the scattering amplitude ${\cal M}$ is a function
of CM energy $\sqrt{s}$ and the scattering angle $\theta$, or
equivalently, of the Mandelstam variables $s$, $t$ and $u$. $p_i, \,
p_f$ are momenta of the incoming and outgoing particles respectively,
where we use the center of mass coordinate system.  ${\cal M}$ for
$V_{L}-V_{L}$ scattering processes for the extended scalar sectors
discussed here can be found in the Appendix~\ref{feynmanamp}.

Partial wave decomposition of the amplitude, followed by application
of the optical theorem,
leads to
\beq\label{optcl}
\sigma=\frac{1}{s}Im\bigg[{\cal
M}(\theta=0)\bigg]=\frac{16\pi}{s}\sum_0^\infty(2l+1)|a_l|^2,
\eeq
which implies
\beq\label{unicond}
|Re(a_l)|\leq\frac12.
\eeq
Here, $a_l$ is the partial wave coefficients corresponding to specific
angular momentum values $l$. This condition is instrumental in
extracting unitarity bounds on any model. At energies large compared
to the gauge boson masses (or, more precisely, for $\sqrt{s} >>
M^2_V$), the equivalence theorem~\cite{Lee:1977eg} implies that
calculations done in terms of the longitudinal modes of the gauge
bosons are same at the lowest order to those using the corresponding
Goldstone bosons, thereby simplifying the calculations. In this limit
the longitudinal gauge boson polarisation vectors can be approximated
as $\epsilon^{\mu}_L(p)\simeq \frac{p^\mu}{M_V}$.

Without a Higgs boson, the unitarity condition is not fulfilled at
high energies.  The inclusion of Higgs-mediated diagrams restores
unitarity rather spectacularly. Any extended scalar sector is in
general expected to satisfy the same requirement, unless one can
come to terms with strongly coupled physics controlling electroweak
interactions at high energy.  Thus the $V_{L}-V_{L}$ scattering
cross-sections in a `well-behaved' new physics scenario should fall at
high centre-of-mass energies. However, if the scattering process
involves the participation of an s-channel resonance at mass $M$, then
one expects a peak at $\sqrt{s} = M$, above which the cross-section
should die down gradually. The energy-dependence of the
cross-sections, along with the appearance (or otherwise) of such
resonant peaks should thus be computed if one has to verify the
imprints of new physics in VBS when the appropriate measurements are
feasible.

We have calculated the amplitudes in different models, using the
exact expressions for polarisation vectors\footnote{One can find the
these expressions of polarisation vectors in
Appendix~\ref{feynmanamp}.}, as we are dealing with the
energy range ($\sim 200 $ GeV $\rightarrow 2000$ GeV). However, we
have checked that at high-energy limit, our results are
consistent with calculations based on the equivalence theorem.

\section{Extended scalar sectors} \label{NewModel}

As has been mentioned already, our purpose is to show the
modifications to the energy-dependence of $V_{L}-V_{L}$ scattering
cross-sections in extended scalar sectors, and point out the
model-dependence in such modifications. Before presenting the results
of our calculation, we outline in the next three sub-sections the
relevant traits of three illustrative used here. In each case,
we re-iterate only those features which influence the calculation of
$V_{L}-V_{L}$ scattering rates. In addition, the obvious constraints
to which each scenario needs to be subjected, such as constraints from
the LHC or precision electroweak data, are mentioned in the
corresponding subsection. We have used parameters consistent with such
constraints in section~\ref{Nemeresult}, and have also ensured that
they do not affect theoretical requirements such as vacuum stability.

\subsection{Type-II two-Higgs doublet model (2HDM)}
In a two Higgs doublet model, an extra $SU(2)_L$ doublet ${\Phi'}$
with hypercharge $Y=1$ is added to the standard model. The extended
scalar potential then looks like~\cite{Branco:2011iw,
  Chakrabarty:2014aya}
\begin{eqnarray}
V(\Phi,\Phi^{\prime}) &=&
m^2_{11}\, \left(\Phi^\dagger \Phi\right)
+ m^2_{22}\, \left(\Phi^{\prime\dagger} \Phi^{\prime}\right) -
 m^2_{12}\, \left(\Phi^\dagger \Phi^{\prime} + \Phi^{\prime\dagger} \Phi\right)
+ \frac{\lambda_1}{2} \left( \Phi^\dagger \Phi \right)^2
+ \frac{\lambda_2}{2} \left( \Phi^{\prime\dagger} \Phi^{\prime} \right)^2
\nonumber \\
&&
+ \lambda_3\, \left(\Phi^\dagger \Phi\,\right) \left(\Phi^{\prime\dagger} \Phi^{\prime}\right)
+ \lambda_4\, \left(\Phi^\dagger \Phi^{\prime}\,\right)\left(\Phi^{\prime\dagger} \Phi\right)
+ \frac{\lambda_5}{2} \left[
\left( \Phi^\dagger\Phi^{\prime} \right)^2
+ \left( \Phi^{\prime\dagger}\Phi \right)^2 \right]
\nonumber \\
&&
+\lambda_6\, \left(\Phi^\dagger \Phi\right)\, \left(\Phi^\dagger\Phi^{\prime} + \Phi^{\prime\dagger}\Phi\right) + \lambda_7\, \left(\Phi^{\prime\dagger} \Phi^{\prime}\right)\, \left(\Phi^\dagger\Phi^{\prime} + \Phi^{\prime\dagger}\Phi\right).
\label{pot}
\end{eqnarray}
As we are not interested in $CP$ violating interactions, we take the
couplings to be real. To avoid tree level FCNCs we will adhere to
Type-II 2HDM scenario in which a discrete symmetry is imposed so that
$\Phi \to -\Phi$, $\Phi^{\prime} \to \Phi^{\prime}$ and
$\psi_{R}^{i}\to -\psi_{R}^{i}$, where $\psi$ stands for charged
leptons or down-type quarks and $i$ represents the generation index.
When this symmetry is exact, $m_{12}$, $\lambda_6$ and $\lambda_7$
vanish. However, to allow a mixing between the two scalar doublets,
the symmetry is softy broken by taking $m_{12}\ne0$. In the Type-II
2HDM, the down-type quarks and the charged leptons couple to $\Phi$
and the up-type quarks couple to
$\Phi^{\prime}$~\cite{Pich:2009sp}. In such a scenario, we have five
massive physical scalars after EWSB --- a pair of charged Higgs
$H^\pm$, two $CP$-even Higgs $h,H$ and one $CP$-odd Higgs $A$. The
mixing angles in the neutral and the charged scalar sectors are
conventionally denoted by $\alpha$ and $\beta$ respectively.

Measurements of the couplings of the SM-like Higgs with the vector bosons
put indirect constraints~\cite{sigstrn} on the models with an extended
scalar sector. For example, a charged Higgs can contribute to
$h\gamma\gamma$ at one loop.  In our analysis, the heavier scalars are
taken to be as heavy as 500~GeV so that $h\gamma\gamma$ constraints
are not that important. $hWW$ and $hZZ$ coupling measurements at
present agrees with SM values, thereby restricting couplings of the
heavier scalars appreciably. As a result, 2HDM is pushed towards the
decoupling regime where couplings of heavier Higgs bosons with SM
particles tend to vanish. We have taken care of all such constraints
at $1\sigma$ in our analysis.

\subsection{Higgs triplet mode (HTM),~$Y=0$ }\label{HTM0}
The scalar sector can be extended by adding a real isospin triplet
$\widetilde{\Phi}$ of hypercharge $Y=0$. The most general scalar
potential is given by~\cite{Chen:2008jg}:
\begin{eqnarray}
V(\Phi,\widetilde{\Phi})&=&\mu_1^2\left(\Phi^\dagger \Phi\right)
+\frac{\mu_2^2} {2} \left(\widetilde{\Phi}^\dagger \widetilde{\Phi}\right)
+\frac{\widetilde{\lambda}_1 }{2} \left(\Phi^\dagger \Phi\right)^2 
+\frac{\widetilde{\lambda}_2}{2} \left(\widetilde{\Phi}^\dagger \widetilde{\Phi}\right)^2 
\nonumber \\ &&
+\frac{\widetilde{\lambda}_3}{2}\left(\Phi^\dagger \Phi\right)^2\left(\widetilde{\Phi}^\dagger \widetilde{\Phi}\right)^2
+\widetilde{\lambda}_4\Phi^\dagger \sigma^a \Phi\widetilde{\Phi}_a\, .
\label{Scalarpot}
\end{eqnarray}
Here, $\mu_1$ and $\mu_2$ are the mass parameters and the coupling
constants $\widetilde{\lambda}_i, ~i=1,4$ are taken to be real. After
EWSB, one is left with the following physical scalar fields: A pair of
charged Higgs $H^\pm$ and two neutral $CP$-even Higgs $h, H$. The
mixing angles corresponding to the charged and neutral scalar sectors
are denoted by $\widetilde{\beta}$ and $\widetilde{\gamma}$
respectively.  $H^{\pm}$ can couple with $W_L^\mp$ and $Z_L$ to
produce a resonance in the $W_L^\mp Z_L\to W_L^\mp Z_L$ channel which
is absent in the Type-II 2HDM.

As HTM models contribute to the $\rho$ parameter at the tree level,
the vacuum expectation value (VEV) of the neutral component of a Higgs
triplet $v_t$ is restricted to be less than
$4$~GeV~\cite{Forshaw:2003kh, Forshaw:2001xq} from measurements of
electroweak precision observables at LEP.

\subsection{Higgs triplet mode (HTM),~$Y=2$ }
There is another variant of the Higgs triplet model with hypercharge
of the triplet $\Delta$ as $Y=2$.  This model has the added virtue
that it can generate neutrino masses. The scalar potential is given
by~\cite{Arhrib:2011uy}:
\begin{eqnarray} 
V(\Phi,\Delta) &=& -m^2_\Phi(\Phi^\dag \Phi)+\frac{\lambda^{\prime}}{4}(\Phi^\dag \Phi)^2+M^2_\Delta {\rm Tr}(\Delta ^\dag \Delta)+ \left(\mu \Phi^{\sf T}i\sigma_2\Delta^\dag\Phi+{\rm h.c.}\right)\,\nonumber\\
&& \lambda^{\prime}_1(\Phi^\dag\Phi){\rm Tr}(\Delta ^\dag \Delta)+\lambda^{\prime}_2\left[{\rm Tr}(\Delta ^\dag \Delta)\right]^2+\lambda^{\prime}_3{\rm Tr}(\Delta ^\dag \Delta)^2+\lambda^{\prime}_4\Phi^\dag\Delta\Delta^\dag\Phi.
\label{eq:Vpd}
\end{eqnarray}
$m_\Phi, M_\Delta~{\rm and}~\mu$ are mass parameters, whereas
$\lambda'$ and $\lambda^{\prime}_i$ ($i=1, 4$) are real coupling
constants. The physical particle spectrum consists of a pair of
doubly-charged Higgs $H^{\pm\pm}$, a pair of singly-charged Higgs
$H^\pm$, two neutral $CP$-even Higgs $h, H$, and a $CP$-odd Higgs
$A$. We have denoted the mixing angles corresponding to the
singly-charged Higgs, $CP$-even neutral Higgs and $CP$-odd neutral
Higgs as $\beta'$, $\gamma'$ and $\delta'$ respectively. $H^{++}$ can
couple with two $W_L^+$ bosons to produce a unique resonance in the
$W_L^+W_L^+\to W_L^+W_L^+$ channel.  Similarly $H^{+}$ can couple with
$W_L^+$ and $Z_L$ to produce a resonance in the $W_L^+Z_L\to W_L^+Z_L$
channel as was in the $Y=0$ triplet model.

In this model, neutrino masses are generated at the tree level. In the
flavour basis, the neutrino mass matrix can be written as
$(M_\nu)_{ij}\propto v_t ~(Y_\nu)_{ij}$, where $Y_\nu$ are the Yukawa
couplings of the Higgs triplet with the neutrinos. Indication of
sub-eV neutrino masses can thus further restrict $v_t$, depending on
the value of $Y_\nu$. For $Y_\nu\sim {\cal O}(1)$, this implies
$v_t\sim{\cal O}(10^{-9})$ GeV. At this limit, the new scalar
particles couple feebly to the SM particles and the decay width of
them would be too small to have a detectable peak at the vector boson
resonances. One can contemplate of the other extreme, when $v_t$ is
saturated to assume its aforesaid maximum value $v_t\lesssim 4$~GeV,
so that $Y_\nu\sim{\cal O}(10^{-9})$. We will work with this extreme
$v_t\sim3$~GeV as this will imply wider resonances in the $V_L V_L$
scattering processes. It will also serve as a conservative choice as
it would mean that the resonances cannot be significantly more wide in
this model.

\section{$V_LV_L$ scattering with extended scalar sectors}\label{Nemeresult}

Next, we demonstrate how it is possible to distinguish among
2HDM, HTM~($Y=0$) and HTM~($Y=2$) using the five VBS processes: $W_L^+
W_L^- \ra W_L^+ W_L^-$, $W_L^+ W_L^- \ra Z_L Z_L$, $Z_L Z_L \ra Z_L
Z_L$, $W_L^+ W_L^+ \ra W_L^+ W_L^+$ and $W_L^+ Z_L\ra W_L^+ Z_L$. One
can immediately see that the mediating scalar can be a neutral
scalar, as also a singly or  doubly charged Higgs.
Thus the very constituents of 2HDM or triplet scenarios
are potential players in the game.

Two things turn out to be crucial here: (a) nature of the energy-dependence,
and (b) the centre-of-mass energy at which the resonances occur.
The shape of the resonance depends on the
decay width, and hence, on the mass and the coupling of the resonating
scalar. Thus an identification of the resonance can guide one
to the theoretical scenario including the particle spectrum.

In any model with an extended scalar sector around a TeV, the very
fact that the $VVh$ interactions ($V \equiv W, Z$ and $h$ = the 125
GeV scalar) are largely SM-like makes the non-resonant additional
contributions small.  In 2HDM, however, these constraints allow enough
parameter space for the heavier scalars to have a large decay width so
that the effect of resonances can be felt for a wider range of
$\sqrt{s}$. In HTM models, however, this is not the case and the
resonances are narrow.  Hence, width of the resonances does not help
in identifying the hypercharge of the scalar triplet.

\begin{figure}[H]
 \begin{center}
 {
 \includegraphics[width=8cm,height=5.85cm, angle=0]{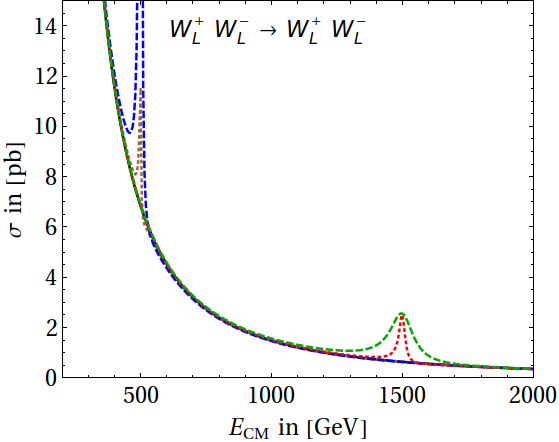}
 \includegraphics[width=8cm,height=6cm, angle=0]{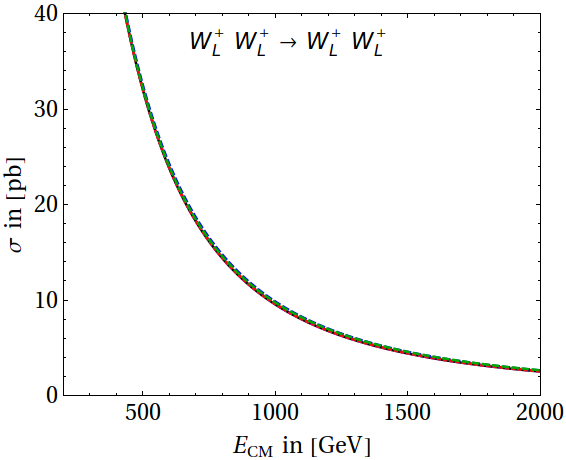}\\
 \includegraphics[width=8cm,height=6cm, angle=0]{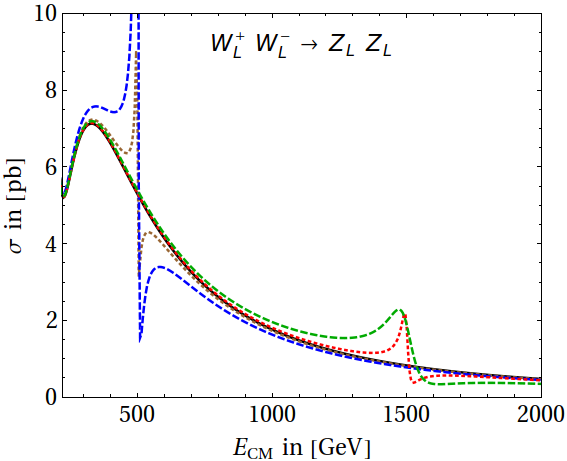}
 \includegraphics[width=8cm,height=6cm, angle=0]{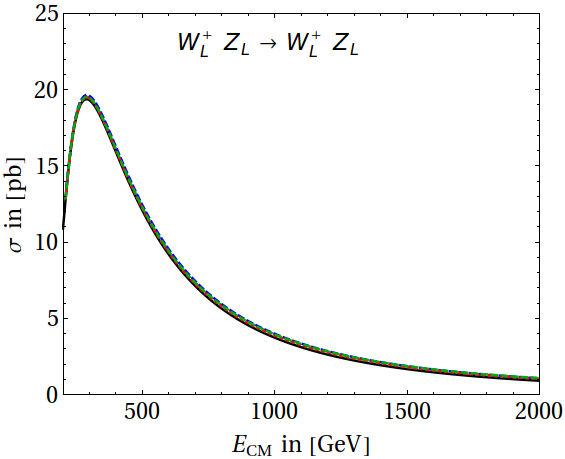} 
\\
\hspace{-2cm} \includegraphics[width=8cm,height=6cm, angle=0]{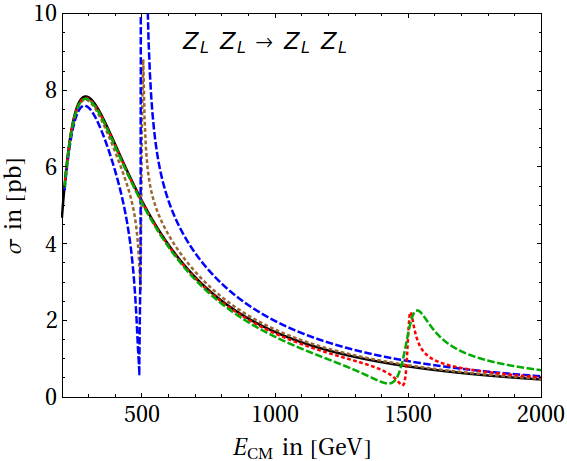}
 \hskip 2pt \includegraphics[width=6cm,height=5cm, angle=0]{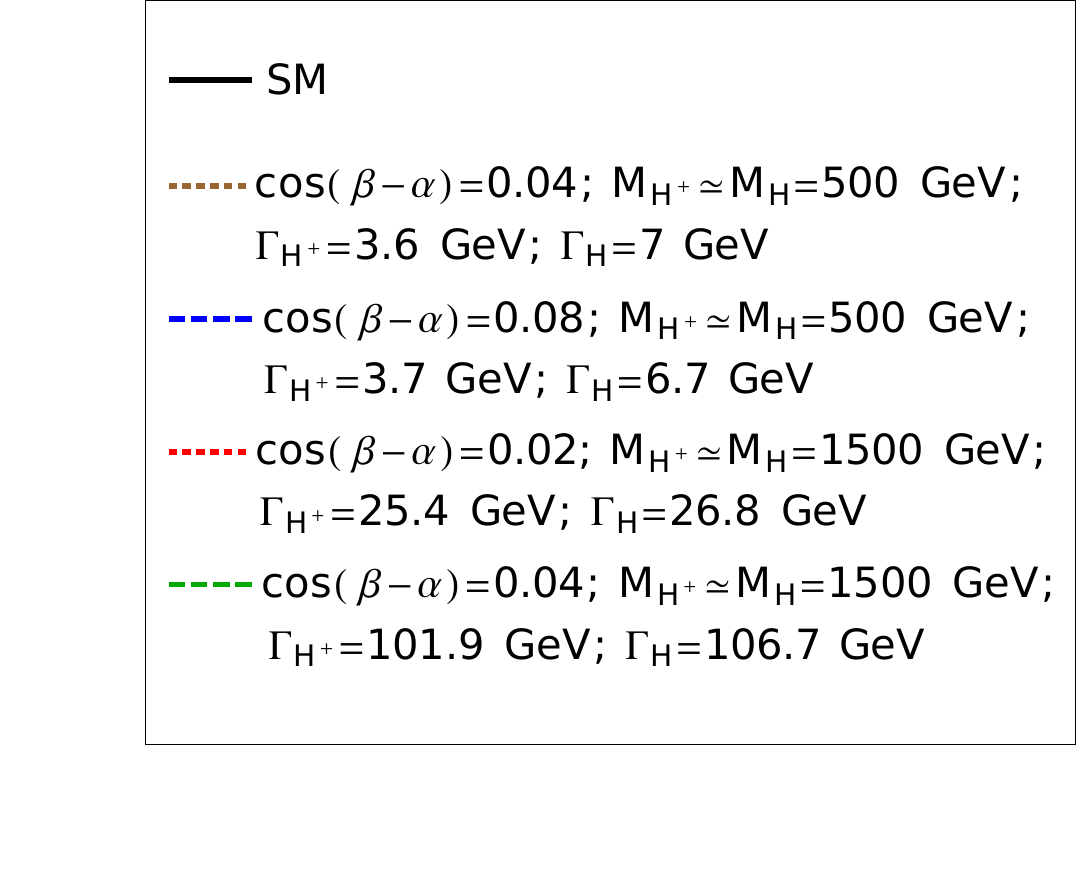}
 \caption{\label{fig:2hdm} \textit{Plots for~$VV$~scattering in 2HDM. } }
 }
 \end{center}
 \end{figure}

 For a 2HDM scenario, the lighter $CP$-even scalar in the particle
 spectrum is usually interpreted as the SM-like Higgs. Going
 especially by the rate of decays into pairs of gauge bosons, the
 couplings of this state is expected to be `nearly SM-like', implying
 that a 2HDM can be feasible largely in the `alignment limit'. Recent
 LHC data are by and large consistent with this limit
 \cite{Broggio:2014mna, Das:2015mwa}. Hence we have performed our analysis almost in
 that limit. Among the five scattering modes, we have resonant peaks
 for only three channels, namely, $W_L^+ W_L^- \ra W_L^+ W_L^-$,
 $W_L^+ W_L^- \ra Z_L Z_L$ and $Z_L Z_L \ra Z_L Z_L$ (see
 Fig.\;\ref{fig:2hdm}) involving the heavier $CP$-even Higgs ($H$). We
 have set its mass ($M_H$) at two benchmark values (500 GeV and
 1500 GeV).  The corresponding decay widths ($\Gamma_H$) can be read
 off from the resonance peaks in Fig.\;\ref{fig:2hdm}. Using the
 high-energy scattering amplitudes given in
 Appendices~\ref{feynmanamp} and \ref{highenergy}, one should be able
 to predict the shapes of plots which contain such resonance peaks. It
 is quite evident from the plots, that apart from the occurrence of
 the peaks, the cross-sections are almost SM-like, as expected in the
 alignment limit. It should also be noted here that in 2HDM, $H^+$
 does not couple to a $W_L^+$ and a $Z_L$. Moreover, $H^{++}$ does not
 exist in this model. As a result there are no resonances in $W_L^+
 W_L^+ \ra W_L^+ W_L^+$ and $W_L^+ Z_L\ra W_L^+ Z_L$ modes.  The
 cross-sections for these processes are also similar to that of SM,
 due to the feeble coupling strength of $H$ with gauge bosons.

\begin{table}[h!]
\begin{center}
    \begin{tabular}{ | c | c | c | c |}
         \hline
     Process &  ${\rm 2HDM}$ & ${\rm HTM}(Y=0)$ & ${\rm HTM}(Y=2)$ \\
    \hline
     $W_L^+ W_L^- \rightarrow W_L^+ W_L^-$ & $\cmark$, $(H)$& $\cmark$, $(H)$&  $\cmark$, $(H)$  \\
     \hline
     $W_L^+ W_L^+ \rightarrow W_L^+ W_L^+$ & $\xmark$ & $\xmark$ &  $\cmark$, $(H^{++})$  \\
     \hline
     $W_L^+ W_L^- \rightarrow Z_L Z_L$  & $\cmark$, $(H)$ & $\cmark$, $(H)$ &  $\cmark$, $(H)$  \\
     \hline
     $W_L^+ Z_L \rightarrow W_L^+ Z_L$  & $\xmark$ & $\cmark$, $(H^+)$ &  $\cmark$, $(H^+)$  \\
     \hline
     $Z_L Z_L \rightarrow Z_L Z_L$  & $\cmark$, $(H)$ & $\cmark$, $(H)$ & $\cmark$, $(H)$ \\
     \hline
     \end{tabular}
    \caption{ \textit{Different scattering processes and corresponding
        mediator particles for resonance in various extended scalar
        sectors. $``\cmark"$ indicates presence of a resonance where
        as $``\xmark"$ corresponds to no resonance peak.} }
    \label{table1}
\end{center}
\end{table}

We have also studied triplet models with two different values of the
$U(1)$ hypercharge ($Y=0$ and 2). In these models we can have
interactions of charged scalars with pairs of gauge bosons. Of these,
we primarily focus on a $Y=2$ HTM. This scenario is relevant in the
context of the Type-II see-saw mechanism of neutrino mass generation,
and it also arises in left-right symmetric gauge theories. Now the
question is, how to isolate such a scenario from a $Y=0$ HTM or even a
Type-II 2HDM models? We summarise our findings in Table~\ref{table1}
which clearly indicates that HTM ($Y=0$) and 2HDM can be distinguished
via a s-channel $H^+$ resonance in $W_L^+ Z_L\ra W_L^+ Z_L$ scattering
process, as $H^+$ couples to $W_L^+$ and $Z_L$ only in the triplet models.
As HTM ($Y=2$) contains a $H^{++}$ that can couple to a pair of $W^+$,
in contrast to the 2HDM and HTM ($Y=0$) models, the distinguishing
feature of this model would be a s-channel $H^{++}$ resonance in
$W_L^+ W_L^+\ra W_L^+ W_L^+$ scattering process.

As mentioned earlier, the shapes of the resonances carry significant
information about the model.  Here, for the sake of illustration we
will concentrate on 2HDM only.  As shown in Fig.\;\ref{fig:2hdm},
$W_L^+ W_L^- \ra W_L^+ W_L^-$ resonances are significantly different
in shape compared to $W_L^+ Z_L\ra W_L^+ Z_L$ and $Z_L Z_L\ra Z_L Z_L$
channel resonances. This is due to the interplay between relative
contributions from the SM and new physics models. The cross-section
can be thought of having three contributions: SM-like, new physics and
an interference between these two. The new physics and the
interference pieces can provide resonance peaks if there exists an
s-channel resonance due to some heavy scalar Higgs. However the
manifestation of such resonances are distinctively different. The new
physics piece for the resonating channel is proportional to
$\frac{M\Gamma}{(s-M^2)^2+(M\Gamma)^2}$, so that it gives a
Breit-Wigner-like contribution. Here $M$ and $\Gamma$ stand for the
mass and the decay width of the heavy scalar, responsible for the
resonance.  In contrary, the corresponding interference piece contains
a factor $\frac{s-M^2}{(s-M^2)^2+(M\Gamma)^2}$, which leads to a shape
that is asymmetric around the pole, which can be understood as
follows. When $\sqrt{s} < M$, there is a destructive interference, if
(otherwise) the relative sign between the SM-like and new physics
piece in the amplitude is positive. On the other hand, for $\sqrt{s} >
M$, the interference is constructive in nature, and hence the
cross-section increases.  If the relative sign between the SM-like and
new physics terms flips, the destructive and then constructive nature
of resonance also gets interchanged.

 Near the pole, the new physics term is more dominant. Away from the
 pole, the interference term may dominate depending on the relative
 sizes of the SM-like and new physics contributions. In each
 resonating VBS mode, the relative sizes are different. For the
 channel $W_L^+ W_L^- \ra W_L^+ W_L^-$ mode, the new physics term
 containing the Breit-Wigner resonance overwhelms the interference piece, 
 so the above-mentioned cross-over
 is not prominent, and one gets a peak.  In the other two resonating modes $W_L^+ Z_L\ra
 W_L^+ Z_L$ and $Z_L Z_L\ra Z_L Z_L$, the
 interference term dominates over the new physics term slightly away
 from the pole, so that one gets a flip. Between $W_L^+ Z_L\ra
 W_L^+ Z_L$ and $Z_L Z_L\ra Z_L Z_L$, the effects of such a cross-over
 are reversed (Fig.~\ref{fig:2hdm}), as the SM contribution flips sign due
 to the absence of a quartic gauge vertex in the latter. Although here 
 we are referring to the 2HDM case, such arguments can also be extended for the triplets.

\begin{figure}[H]
 \begin{center}
 {
 \includegraphics[width=8cm,height=5.85cm, angle=0]{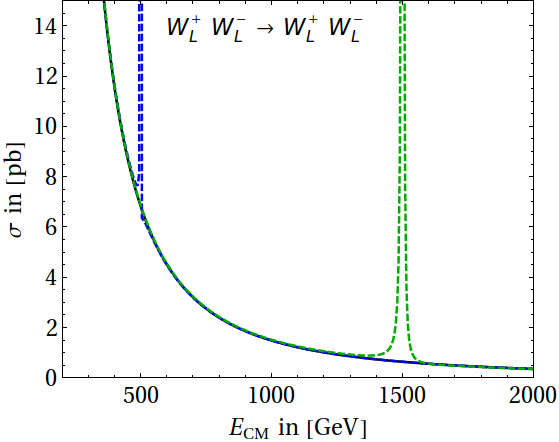}
 \includegraphics[width=8cm,height=6cm, angle=0]{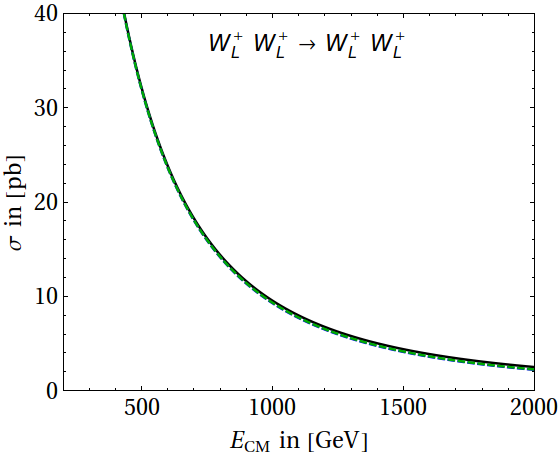}\\
 \includegraphics[width=8cm,height=6cm, angle=0]{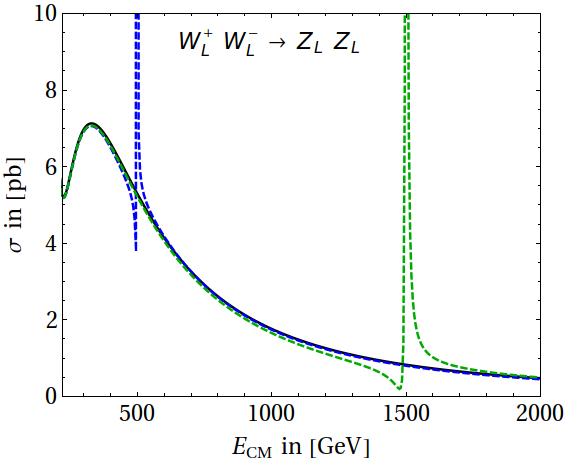}
 \includegraphics[width=8cm,height=6cm, angle=0]{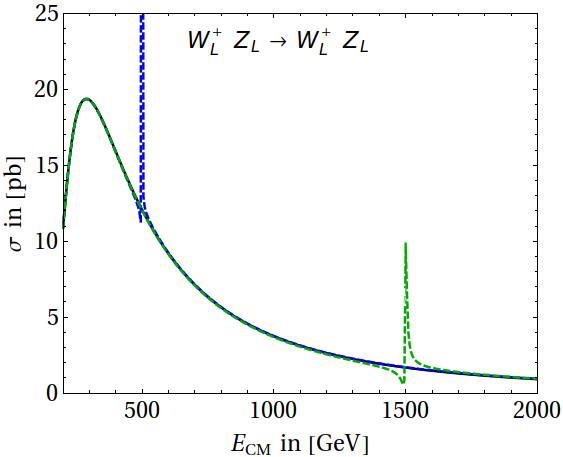}
\\
\hspace{-2cm}
\includegraphics[width=8cm,height=6cm, angle=0]{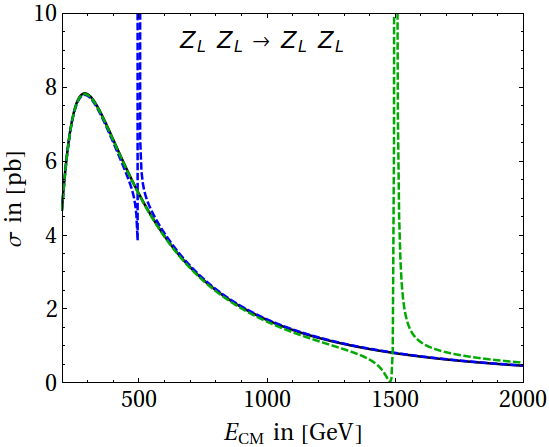}
\hskip 2pt
\includegraphics[width=6cm,height=5cm, angle=0]
{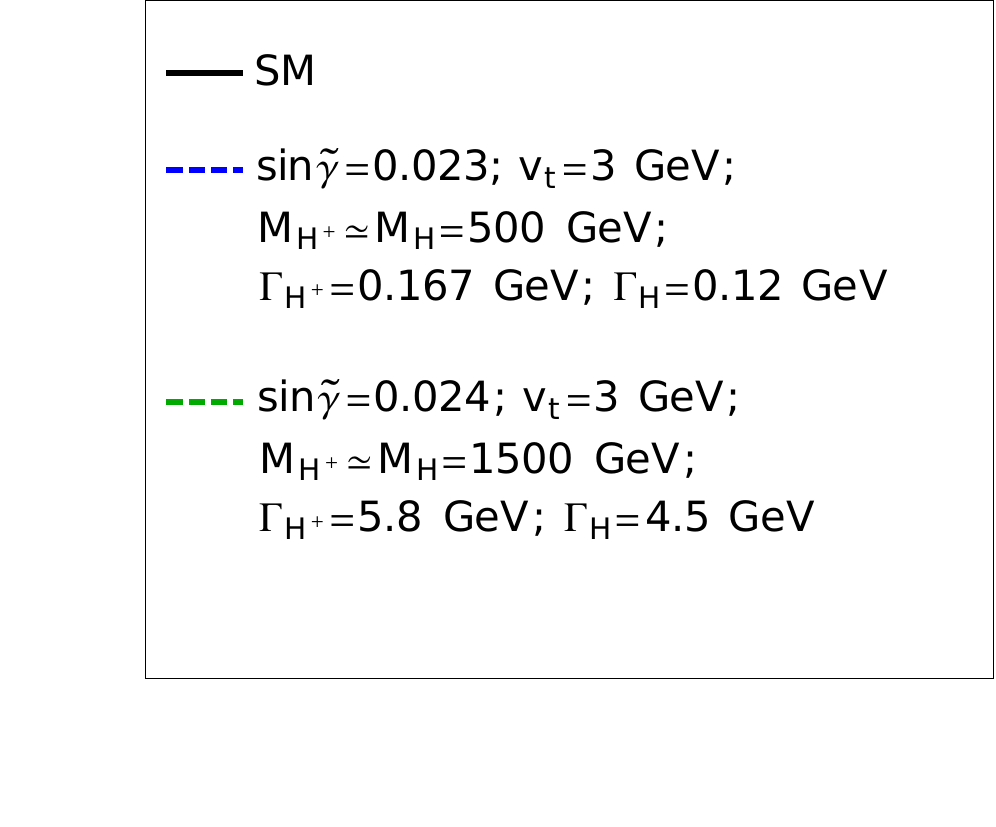} 
 \caption{\label{fig:htm0} \textit{Plots for~$VV$~scattering in Y=0 HTM. } }
}
 \end{center}
 \end{figure}

\begin{figure}[H]
\begin{center}
{	
 \includegraphics[width=8cm,height=6cm, angle=0]{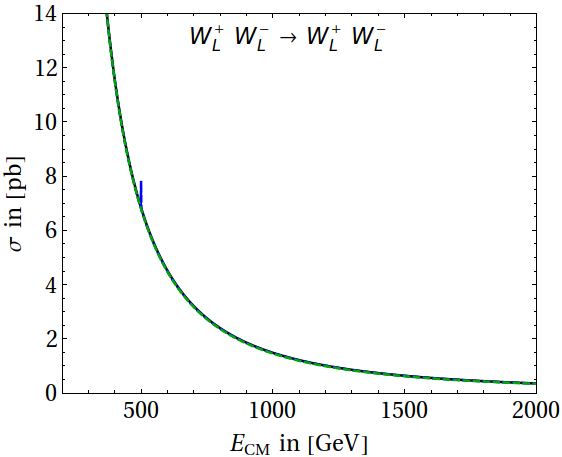}
 \includegraphics[width=8cm,height=6cm, angle=0]{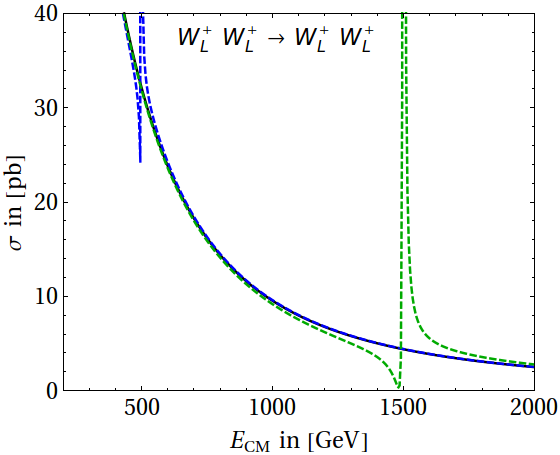}
 \includegraphics[width=8cm,height=6cm, angle=0]{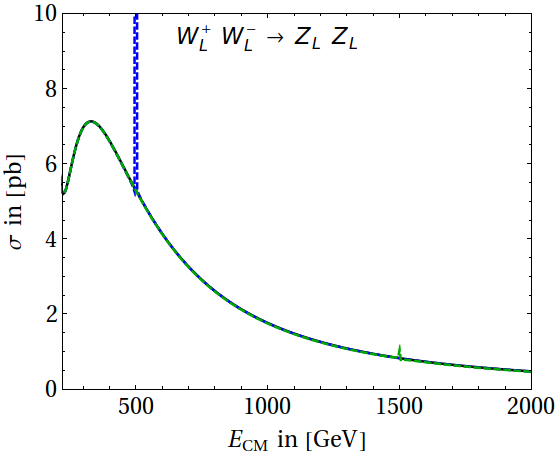}
 \includegraphics[width=8cm,height=6cm, angle=0]{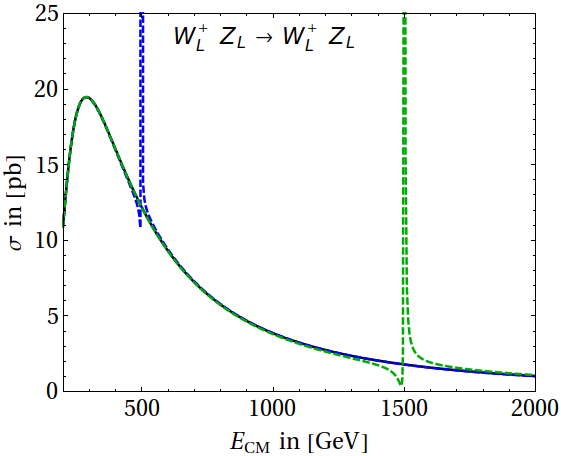}  
\\
\hspace{-2cm}
 \includegraphics[width=8cm,height=6cm, angle=0]{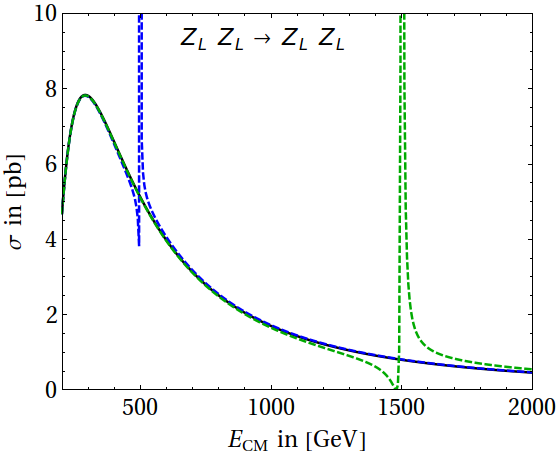}	
\hskip 2pt \includegraphics[width=6cm,height=5cm, angle=0]{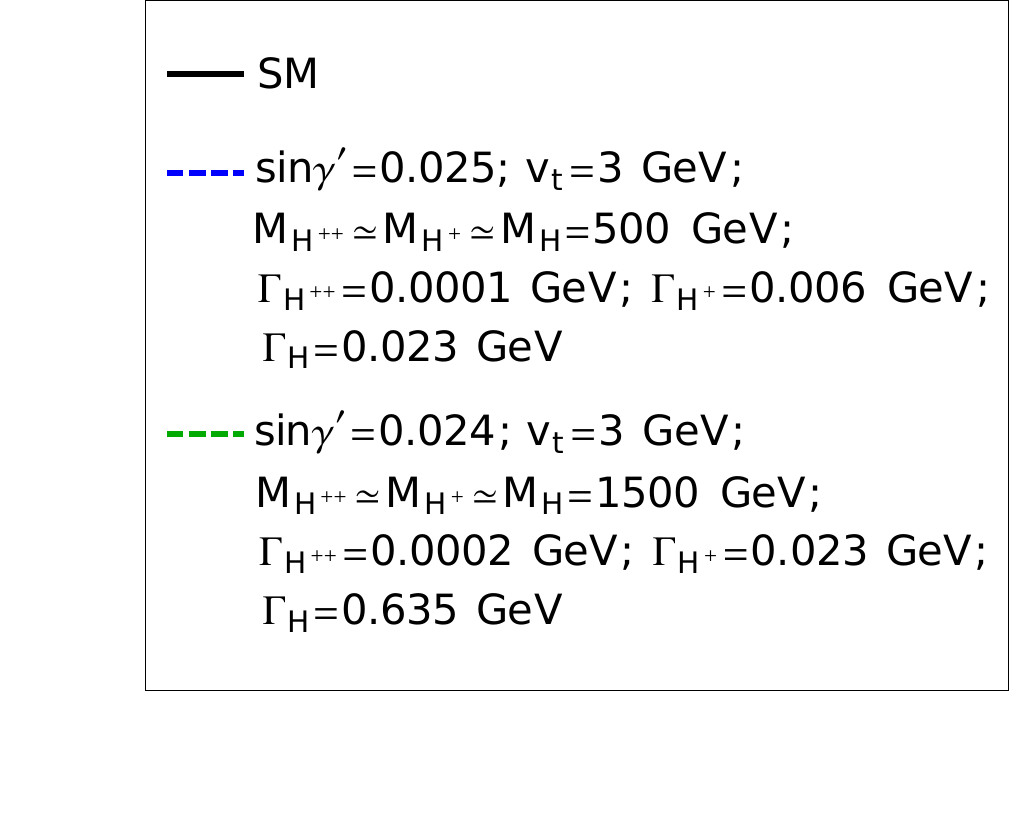}
\caption{\label{fig:htm2} \textit{Plots for~$VV$~scattering in Y=2 HTM. } }
 }
 \end{center}
 \end{figure}

Due to the aforementioned interference between the SM and the new physics contributions, the manifestations of resonances are of different nature than a simple Breit-Wigner one. For example, 
the CM energy at which the peak or the flip occurs can be shifted from the mass of the resonating scalar and this shift depends not only on the width of the scalar but also on the new physics parameters. As an illustration of this, in 2HDM, for the chosen benchmark points, one can experience such shifts: In Fig.~\ref{fig:2hdm} for a resonating scalar mass of 1500\,GeV, the peak shifts by $\sim 20$\,GeV, whereas for triplets (Figs.~\ref{fig:htm0}, \ref{fig:htm2}) such shifts are rather tiny $\sim1-2$\,GeV.  As the magnitude of the triplet VEV is  severely restricted by the $\rho$-parameter,  the decay width of the scalar, and hence the shift of the peak or flip from the resonating scalar mass are rather small compared to what could come from the doublet scalars.

One should be similarly careful in interpreting the width of the resonances as it depends not only on the decay width of the resonating scalar particle but also 
directly on the new physics parameters.

 In the high-energy limit ($E_{CM} \gg M_H$), the amplitudes can be
 expressed as a power series in the energy (see
 Eq.\;\ref{energyseries}). In this limit the terms proportional
to  $E^4_{CM}$ and $E^2_{CM}$ of the amplitude become zero. The
 remaining terms are either independent of energy or go in negative
 powers of energy, so that the cross-section decreases with rising
 energy, thus ensuring perturbative unitarity.  We present analytical
 expressions for the dominant terms at such energies in
 Appendix~\ref{highenergy} to help the reader in reproducing
 cross-sections at a very high $\sqrt{s}$.

\section{Summary and conclusion}\label{conclusionSUM}
For the scattering of longitudinally polarised weak gauge bosons, 
we point out that the resonances arising in some extended scalar
sectors can be noticed in the distribution in the subprocess
centre-of-mass energies, which can play a complementary (or
confirmatory) role in the search of such new scalars. Both the new
physics term containing a Breit-Wigner-resonance and the interference
term between the new physics and the standard model contributions
involved in computing cross-sections conspire to demonstrate the
effect of new scalar resonances.

We have presented our results on different scattering channels for
longitudinally polarised gauge bosons in various extended scalar
sectors. VBS processes are sensitive to the scalar sector as the
scalars ensure the unitarity of the scattering matrix. 

We worked with unapproximated polarisation vectors and used complete set
of Feynman diagrams. This is the reason we have provided the complete
analytical expressions which can help the reader to compute the
discussed VBS processes in various extensions of the scalar sector of
the standard model. Exact forms of the polarisation vectors have been used,
and all relevant Feynman diagrams as well as the complete analytical expressions 
have been presented in the Appendix.

For the sake of illustration, we have chosen three models with an
extended scalar sector, namely the 2HDM and triplet models with
hypercharge $Y=0$ and $2$. All these models are endowed with heavier
scalars -- for the chosen benchmark points the masses are at 500~GeV
or at 1500~GeV. From the presence of resonances at various VBS modes,
we have tried to identify the underlying model. 

The shape of the invariant mass distribution in
the vicinity of the resonance is found to offer rather distinctive features in this respect.  First, the decay width of the mediating scalar, occurring in the Breit-Wigner propagator, along with the other parameters of the theory, determines the small shift of the resonant point from the actual mass of the charged or neutral scalar involved. This in turn depends on the scalar VEV, and thus provides a substantial distinction between the cases where a member of an SU(2) doublet or a triplet is the mediator, because the VEV of the latter is constrained to be much smaller. In addition, the model parameters (especially the VEVs) also determine the relative contributions of  the purely `new physics part' and its interference term with the SM in the squared matrix element. When the former is bigger, one usually observes just a resonant peak. With sizeable or dominant contribution from the latter,  a sign flip in the propagator causes a dip followed by a peak (or the other way around) in the plot against $\sqrt{s}$.

This work is aimed at pointing out the above model-specific features in longitudinal gauge boson scattering, which can supplement the phenomenology involving the scalar sector itself.   A more detailed description of the collider observables that enable one to extract faithful information on the resonant peaks (or accompanying dips) will be presented in a later study.

\section{Acknowledgements}
The work of N.K. is supported by a fellowship from University Grants
Commission, India. B.M. and A.S. acknowledge the funding available
from the Department of Atomic Energy, Government of India, for the
Regional Centre for Accelerator based Particle Physics~(RECAPP),
Harish-Chandra Research Institute.  S.R. is funded by the Department
of Science and Technology, India via Grant No. EMR/2014/001177. 
The visit of A.S. at IIT Indore was also supported from this grant.  
S.R. acknowledges hospitality of RECAPP while this work was in progress.


\newpage
\begin{appendices}
\renewcommand{\thesection}{\Alph{section}}
\renewcommand{\theequation}{\thesection-\arabic{equation}} 
\allowdisplaybreaks
\setcounter{equation}{0}  
\section{Amplitudes for individual diagrams in the different modes of $V_L V_L$ scattering}\label{feynmanamp}

Let $p_1,~p_2$ are the four momenta of initial state gauge bosons and $k_1,~k_2$ are that for the final state gauge bosons respectively. 
\begin{itemize}
\item Let, $\epsilon(p)$ be the polarization four vector of a gauge bosons $V(\equiv W^\pm,~Z$) with four momentum $p$ which satisfies the relation $\epsilon(p)\cdot p=0$.
It can be written as, $\epsilon_\mu(p)\equiv\{\frac{|{\bf p}|}{M_V},\frac{E_V}{M_V}\hat{p}\}$, where $E_V$ is the energy of the gauge boson and is given by $E_V=\sqrt{|{\bf p}|^2+M_V^2}$. Here, $M_V$ is the mass of $V$.
For illustrative purpose, we choose the notations as $ \epsilon_1 \equiv  \epsilon(p_1)$, $\epsilon_2\equiv \epsilon(p_2) $, $\epsilon_3 \equiv \epsilon(k_1) $ and $\epsilon_4\equiv \epsilon(k_2)$.
\item Mandelstam variables are defined as: $s=(p_1+p_2)^2$; $t=(p_1-k_1)^2$; $u=(p_1-k_2)^2$. In the following, $x \equiv \cos\theta$, where $\theta$ is the scattering angle.
\item We use the shorthand notations $c_W\equiv\cos{\theta_W} $  and  $ s_W\equiv\sin{\theta_W}$, where $\theta_W$ is the Weinberg angle.
\end{itemize}

In the following we present the contributing Feynman diagrams and scattering amplitudes for various vector boson scattering modes in terms of total energy in the centre of momentum frame $E_{CM} \equiv \sqrt{s}$ and scattering angle. We use Unitary gauge in our calculations. Note that the coefficients $C$, $C^{\prime}$, $\hat{C}$, $\tilde{C}$ are model dependent and can be found in Appendix C. 

\subsection{$W_L^+(p_1)~+~ W_L^-(p_2) \rightarrow W_L^+(k_1)~+~W_L^-(k_2)$}
 \begin{figure}[h!]
 \begin{center}
 {
 \includegraphics[width=16cm,height=10cm, angle=0]{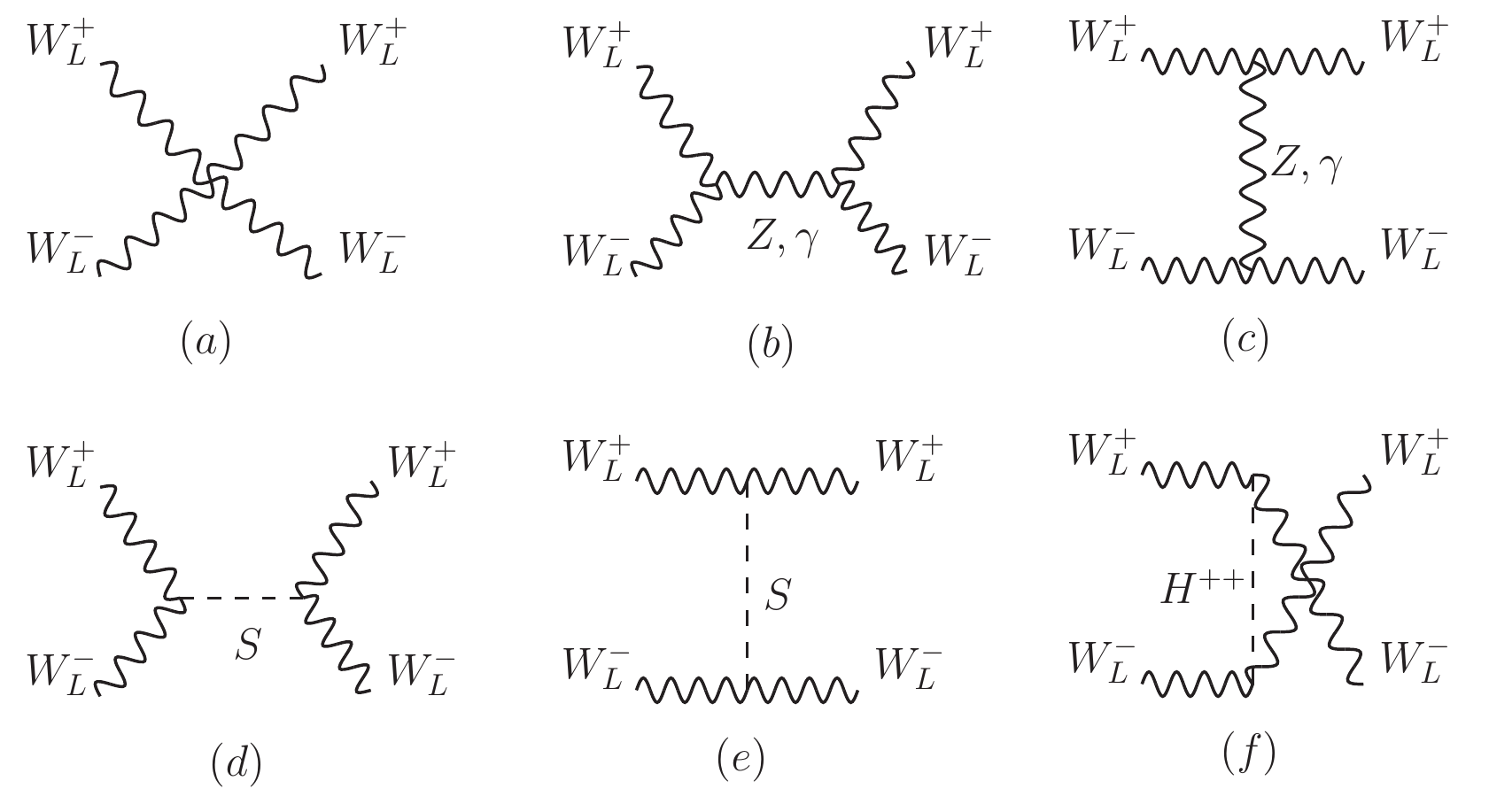}
 \caption{\label{fig:WpWmWpWm} \textit{Generic Feynman diagrams for~~$W_L^+(p_1)~+~ W_L^-(p_2) \rightarrow W_L^+(k_1)~+~W_L^-(k_2)$~~scattering.} }
 }
 \end{center}
 \end{figure}
\bea
(a)& {\cal M}_{\rm p}&\hspace{-2ex} =~~\frac{E_{CM}^2 g_2^2}{16 M_{W}^4} \bigg\{8 M_{W}^2 (1-3 x)-E_{CM}^2 \left(3-6 x-x^2\right)\bigg\}.\nn\\
(b)& {\cal M}^{\gamma+{Z}}_{s} &\hspace{-2ex} =~ -\frac{g_2^2}{4 M_{W}^4} \left( \frac{s_W^2}{s} + \frac{c_W^2}{s-M_Z^2}\right) \bigg( E_{CM}^6-12 E_{CM}^2 M_{W}^4-16 M_{W}^6\bigg) x.\nn\\
(c)&{\cal M}^{\gamma+{Z}}_{t} &\hspace{-2ex} =~ -\frac{g_2^2}{32M_{W}^4} \left( \frac{s_W^2}{t} + \frac{c_W^2}{t-M_Z^2}\right)\bigg\{-64 M_{W}^6 (1+x)+E_{CM}^6 (1-x)^2 (3+x)\nn \\&&\hspace{2cm}+16 E_{CM}^2 M_{W}^4 \left(1-7 x+10 x^2\right)-4 E_{CM}^4 M_{W}^2 \left(3-13 x+9 x^2+x^3\right)\bigg\}.\nn\\
(d)& {\cal M}^{S}_{s} &\hspace{-2ex} =~  - \frac{(C~g_2 M_W)^2}{s-M_S^2}~\frac{\left(E_{CM}^2-2 M_{W}^2\right)^2}{4 M_W^4}.\nn\\
(e)& {\cal M}^{S}_{t} &\hspace{-2ex} =~  - \frac{(C~g_2 M_W)^2}{t-M_S^2}~\frac{\{4 M_W^2+E_{CM}^2 (1-x)\}^2}{16 M_W^4}.\nn\\
(f)&{\cal M}^{H^{++}}_{u} &\hspace{-2ex} =~  - \frac{(\hat{C}~g_2 M_W)^2}{u-M^2_{H^{++}}}~\frac{\{-4 M_W^2+E_{CM}^2 (1+x)\}^2}{16 M_W^4}.\nn
\eea


\subsection{$W_L^+(p_1)~+~ W_L^+(p_2) \rightarrow W_L^+(k_1)~+~W_L^+(k_2)$}
 \begin{figure}[h!]
 \begin{center}
 {
 \includegraphics[width=16cm,height=10cm, angle=0]{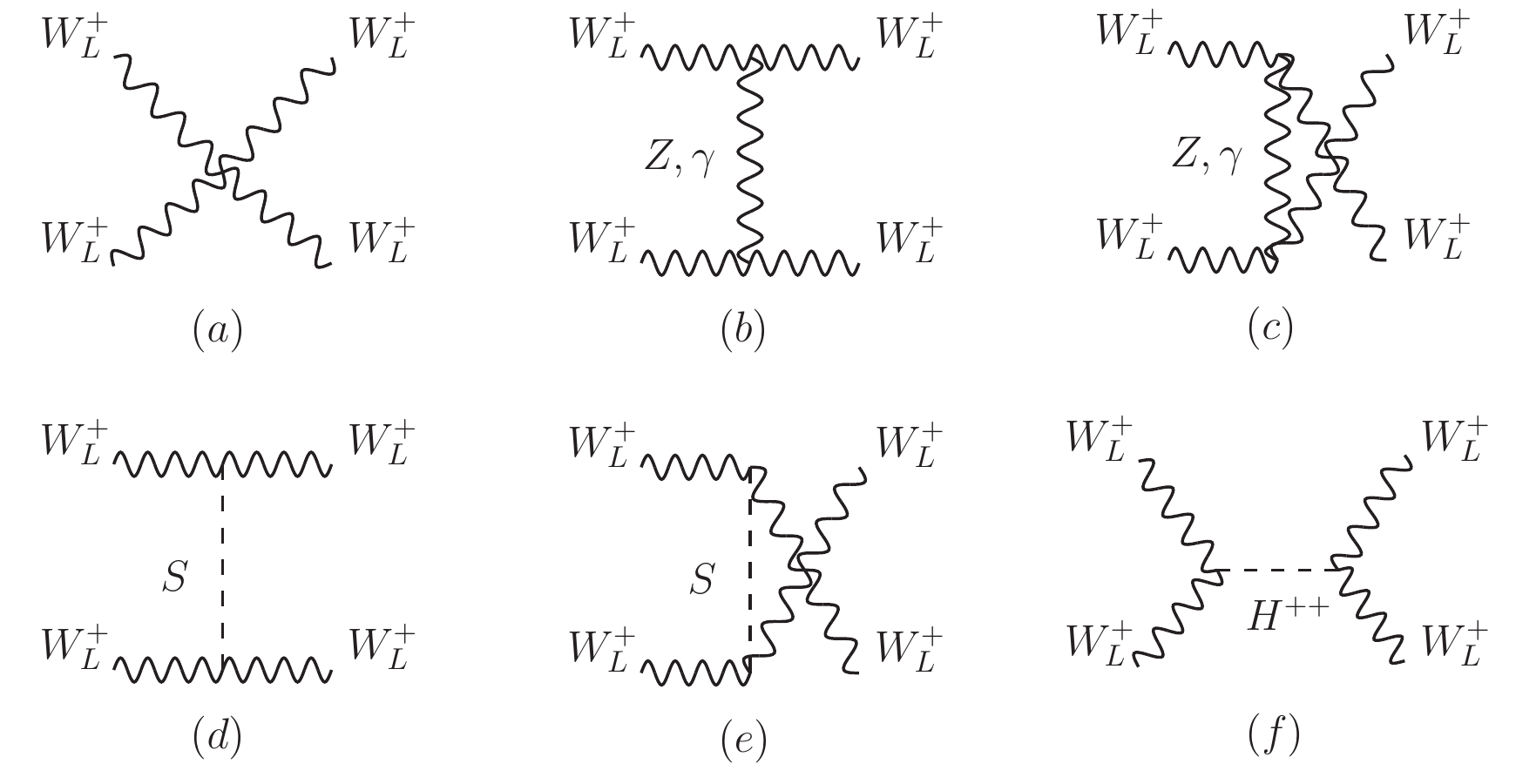}
 \caption{\label{fig:WpWpWpWp} \textit{Generic Feynman diagrams for~~$W_L^+(p_1)~+~ W_L^+(p_2) \rightarrow W_L^+(k_1)~+~W_L^+(k_2)$~~scattering. } }
 }
 \end{center}
 \end{figure}
\bea
(a)& {\cal M}_{\rm p} &\hspace{-2ex} =~~  \frac{E_{CM}^2 g_2^2 c_W^2}{8M_{W}^2 M_{Z}^2}\bigg\{4 \left(M_{W}^2+M_{Z}^2\right)-E_{CM}^2 \left(3-x^2\right)\bigg\}.\nn\\
(b)& {\cal M}^{W}_{t} &\hspace{-2ex} =~ -\frac{ c_W^2 g_2^2}{32M_{W}^4 M_{Z}^2(t-M_W^2)} \bigg\{  3 E_{CM}^6 M_{W}^2-4 E_{CM}^4 M_{W}^4-10 E_{CM}^4 M_{W}^2 M_{Z}^2\nn\\
&&\hspace{2cm}+8 E_{CM}^2 M_{W}^4 M_{Z}^2+2 E_{CM}^4 M_{Z}^4+16 E_{CM}^2 M_{W}^2 M_{Z}^4-96 M_{W}^4 M_{Z}^4\nn\\
&&\hspace{2cm}-8 E_{CM}^2 M_{Z}^6+32 M_{W}^2 M_{Z}^6+x (E_{CM}^4 M_{W}^2\text{ }x^2-5 E_{CM}^4 M_{W}^2\nn\\
&&\hspace{2cm}+12 E_{CM}^2 M_{W}^4+24 E_{CM}^2 M_{W}^2 M_{Z}^2+16 M_{W}^4 M_{Z}^2\nn\\
&&\hspace{2cm}-4 E_{CM}^2 M_{Z}^4)\sqrt{(E_{CM}^2-4 M_{W}^2) (E_{CM}^2-4 M_{Z}^2)}+E_{CM}^6 M_{W}^2 x^2\nn\\
&&\hspace{2cm}-16 E_{CM}^4 M_{W}^4 x^2+32 E_{CM}^2 M_{W}^6 x^2-22 E_{CM}^4 M_{W}^2 M_{Z}^2 x^2\nn\\
&&\hspace{2cm}+96 E_{CM}^2 M_{W}^4 M_{Z}^2 x^2+2 E_{CM}^4 M_{Z}^4 x^2+32 E_{CM}^2 M_{W}^2 M_{Z}^4 x^2 \bigg\}.\nn\\
(c)&{\cal M}^{W}_{u} &\hspace{-2ex} =~  -\frac{ c_W^2 g_2^2}{32M_{W}^4 M_{Z}^2(u-M_W^2)} \bigg\{  3 E_{CM}^6 M_{W}^2-4 E_{CM}^4 M_{W}^4-10 E_{CM}^4 M_{W}^2 M_{Z}^2
\nn\\
&&\hspace{2cm}
+8 E_{CM}^2 M_{W}^4 M_{Z}^2+2 E_{CM}^4 M_{Z}^4+16 E_{CM}^2 M_{W}^2 M_{Z}^4-96 M_{W}^4 M_{Z}^4\nn\\
&&\hspace{2cm}-8 E_{CM}^2 M_{Z}^6+32 M_{W}^2 M_{Z}^6- x (E_{CM}^4 M_{W}^2\text{  }x^2-5 E_{CM}^4 M_{W}^2\nn\\
&&\hspace{2cm}+12 E_{CM}^2 M_{W}^4+24 E_{CM}^2 M_{W}^2 M_{Z}^2+16 M_{W}^4 M_{Z}^2\nn\\
&&\hspace{2cm}-4 E_{CM}^2 M_{Z}^4)\sqrt{(E_{CM}^2-4 M_{W}^2) (E_{CM}^2-4 M_{Z}^2)}+E_{CM}^6 M_{W}^2 x^2\nn\\
&&\hspace{2cm}-16 E_{CM}^4 M_{W}^4 x^2+32 E_{CM}^2 M_{W}^6 x^2-22 E_{CM}^4 M_{W}^2 M_{Z}^2 x^2\nn\\
&&\hspace{2cm}+96 E_{CM}^2 M_{W}^4 M_{Z}^2 x^2+2 E_{CM}^4 M_{Z}^4 x^2+32 E_{CM}^2 M_{W}^2 M_{Z}^4 x^2 \bigg\}.\nn\\
(d)& {\cal M}^{S}_{s} &\hspace{-2ex} =~  -\frac{(C~g_2 M_W) (C^{\prime}~\frac{g_2 M_Z}{c_W})}{s-M_S^2}~ \frac{\left(E_{CM}^2-2 M_W^2\right) \left(E_{CM}^2-2 M_Z^2\right)}{4 M_W^2 M_Z^2}.\nn\\
(e)& {\cal M}^{H^+}_{t} &\hspace{-2ex} =~  -\frac{(\widetilde{C}~\frac{g_2 M_Z}{c_W})^2}{t-M_{H^+}^2}~ \frac{ \big \{ \sqrt{\left(E_{CM}^2-4 M_W^2\right) \left(E_{CM}^2-4 M_Z^2\right)}-E_{CM}^2 x\big \}^2}{16 M_W^2 M_Z^2}.\nn\\
(f)& {\cal M}^{H^+}_{u} &\hspace{-2ex} =~  -\frac{(\widetilde{C}~\frac{g_2 M_Z}{c_W})^2}{u-M_{H^+}^2} ~\frac{\big \{ \sqrt{\left(E_{CM}^2-4 M_W^2\right) \left(E_{CM}^2-4 M_Z^2\right)}+E_{CM}^2 x\big\}^2}{16 M_W^2 M_Z^2}.\nn
\eea

\subsection{$W_L^+(p_1)~+~ W_L^-(p_2) \rightarrow Z_L(k_1)~+~Z_L(k_2)$}
 \begin{figure}[h!]
 \begin{center}
 {
 \includegraphics[width=16cm,height=10cm, angle=0]{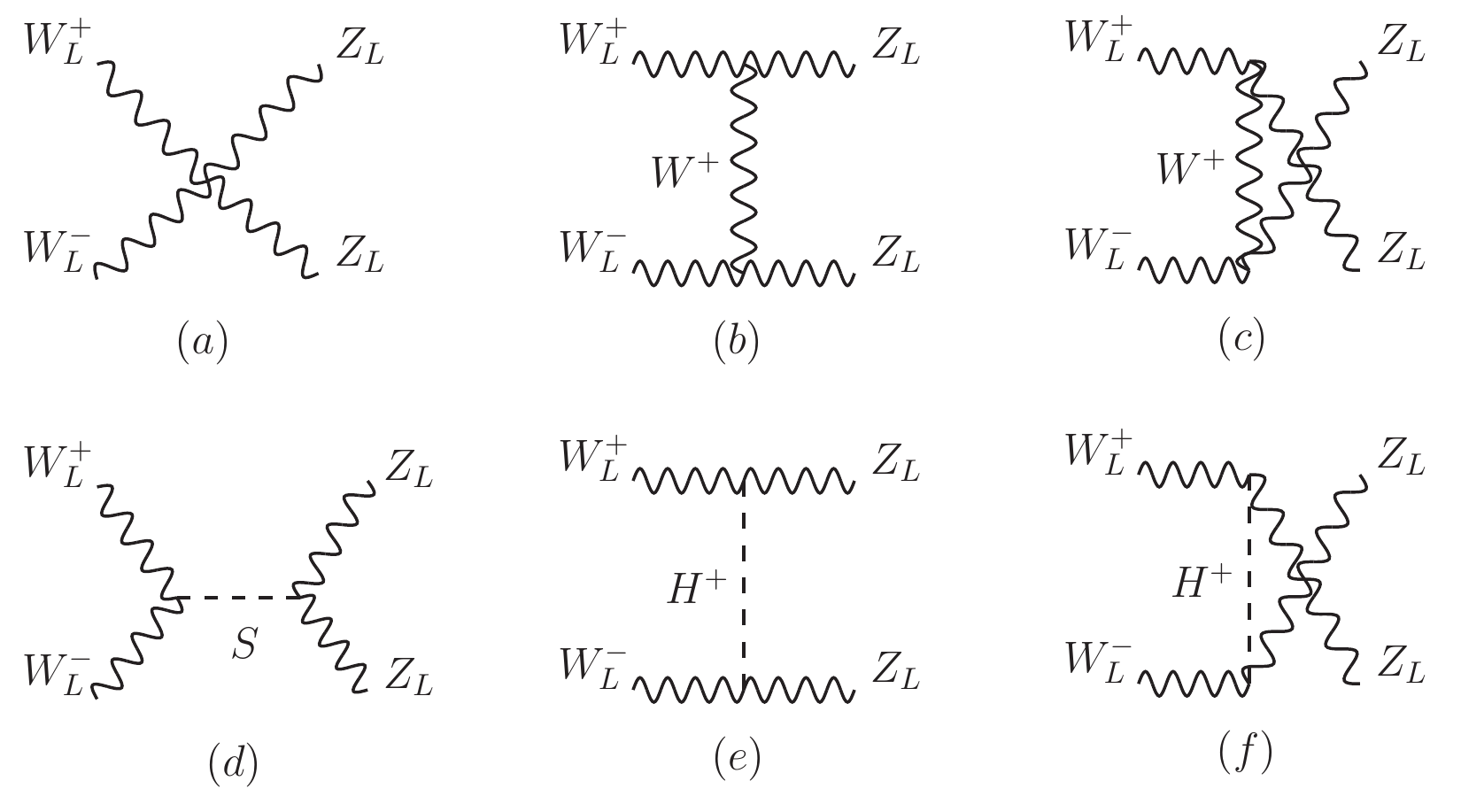}
 \caption{\label{fig:WpWmZZ} \textit{Generic Feynman diagrams  for~~$W_L^+(p_1)~+~ W_L^-(p_2) \rightarrow Z_L(k_1)~+~Z_L(k_2)$~~scattering. } }
 }
 \end{center}
 \end{figure}
\bea
(a)& {\cal M}_{\rm p} &\hspace{-2ex}=~ \frac{E_{CM}^2 g_2^2 c_W^2}{8M_{W}^2 M_{Z}^2}\bigg\{4 \left(M_{W}^2+M_{Z}^2\right)-E_{CM}^2 \left(3-x^2\right)\bigg \}.\nn\\
(b)&  {\cal M}^{W}_{t} &\hspace{-2ex}=~ -\frac{ c_W^2 g_2^2}{32M_{W}^4 M_{Z}^2(t-M_W^2)} \bigg\{  3 E_{CM}^6 M_{W}^2-4 E_{CM}^4 M_{W}^4-10 E_{CM}^4 M_{W}^2 M_{Z}^2\nn\\
&&\hspace{2cm}+8 E_{CM}^2 M_{W}^4 M_{Z}^2+2 E_{CM}^4 M_{Z}^4+16 E_{CM}^2 M_{W}^2 M_{Z}^4-96 M_{W}^4 M_{Z}^4-8 E_{CM}^2 M_{Z}^6\nn\\
&&\hspace{2cm}+32 M_{W}^2 M_{Z}^6+x (E_{CM}^4 M_{W}^2\text{  }x^2-5 E_{CM}^4 M_{W}^2+12 E_{CM}^2 M_{W}^4-4 E_{CM}^2 M_{Z}^4\nn\\
&&\hspace{2cm}+16 M_{W}^4 M_{Z}^2 +24 E_{CM}^2 M_{W}^2 M_{Z}^2 )\sqrt{(E_{CM}^2-4 M_{W}^2) (E_{CM}^2-4 M_{Z}^2)}\nn\\
&&\hspace{2cm}+E_{CM}^6 M_{W}^2 x^2-16 E_{CM}^4 M_{W}^4 x^2+32 E_{CM}^2 M_{W}^6 x^2-22 E_{CM}^4 M_{W}^2 M_{Z}^2 x^2\nn\\
&&\hspace{2cm}+96 E_{CM}^2 M_{W}^4 M_{Z}^2 x^2+2 E_{CM}^4 M_{Z}^4 x^2+32 E_{CM}^2 M_{W}^2 M_{Z}^4 x^2 \bigg\}.\nn\\
(c)& {\cal M}^{W}_{u} &\hspace{-2ex}=~ -\frac{ c_W^2 g_2^2}{32M_{W}^4 M_{Z}^2(u-M_W^2)} \bigg\{  3 E_{CM}^6 M_{W}^2-4 E_{CM}^4 M_{W}^4-10 E_{CM}^4 M_{W}^2 M_{Z}^2\nn\\
&&\hspace{2cm}+8 E_{CM}^2 M_{W}^4 M_{Z}^2+2 E_{CM}^4 M_{Z}^4+16 E_{CM}^2 M_{W}^2 M_{Z}^4-96 M_{W}^4 M_{Z}^4-8 E_{CM}^2 M_{Z}^6\nn\\
&&\hspace{2cm}+32 M_{W}^2 M_{Z}^6- x (E_{CM}^4 M_{W}^2\text{  }x^2-5 E_{CM}^4 M_{W}^2+12 E_{CM}^2 M_{W}^4 -4 E_{CM}^2 M_{Z}^4\nn\\
&&\hspace{2cm}+16 M_{W}^4 M_{Z}^2 +24 E_{CM}^2 M_{W}^2 M_{Z}^2  )\sqrt{(E_{CM}^2-4 M_{W}^2) (E_{CM}^2-4 M_{Z}^2)}\nn\\
&&\hspace{2cm}+E_{CM}^6 M_{W}^2 x^2-16 E_{CM}^4 M_{W}^4 x^2+32 E_{CM}^2 M_{W}^6 x^2-22 E_{CM}^4 M_{W}^2 M_{Z}^2 x^2\nn\\
&&\hspace{2cm}+96 E_{CM}^2 M_{W}^4 M_{Z}^2 x^2+2 E_{CM}^4 M_{Z}^4 x^2+32 E_{CM}^2 M_{W}^2 M_{Z}^4 x^2 \bigg \}.\nn\\
(d)& {\cal M}^{S}_{s} &\hspace{-2ex}=~ -\frac{(C~g_2 M_W) (C^{\prime}~\frac{g_2 M_Z}{c_W})}{s-M_S^2}~ \frac{\left(E_{CM}^2-2 M_W^2\right) \left(E_{CM}^2-2 M_Z^2\right)}{4 M_W^2 M_Z^2}.\nn\\
(e)& {\cal M}^{H^+}_{t} &\hspace{-2ex}=~ -\frac{(\widetilde{C}~\frac{g_2 M_Z}{c_W})^2}{t-M_{H^+}^2}~ \frac{\big\{\sqrt{\left(E_{CM}^2-4 M_W^2\right) \left(E_{CM}^2-4 M_Z^2\right)}-E_{CM}^2 x\big\}^2}{16 M_W^2 M_Z^2}.\nn\\
(f)& {\cal M}^{H^+}_{u} &\hspace{-2ex}=~ -\frac{(\widetilde{C}~\frac{g_2 M_Z}{c_W})^2}{u-M_{H^+}^2} ~\frac{\big\{ \sqrt{\left(E_{CM}^2-4 M_W^2\right) \left(E_{CM}^2-4 M_Z^2\right)}+E_{CM}^2 x\big\}^2}{16 M_W^2 M_Z^2}.\nn
\eea

\subsection{$W_L^+(p_1)~+~ Z_L(p_2) \rightarrow W_L^+(k_1)~+~Z_L(k_2)$}
 \begin{figure}[h!]
 \begin{center}
 {
\includegraphics[width=16cm,height=10cm, angle=0]{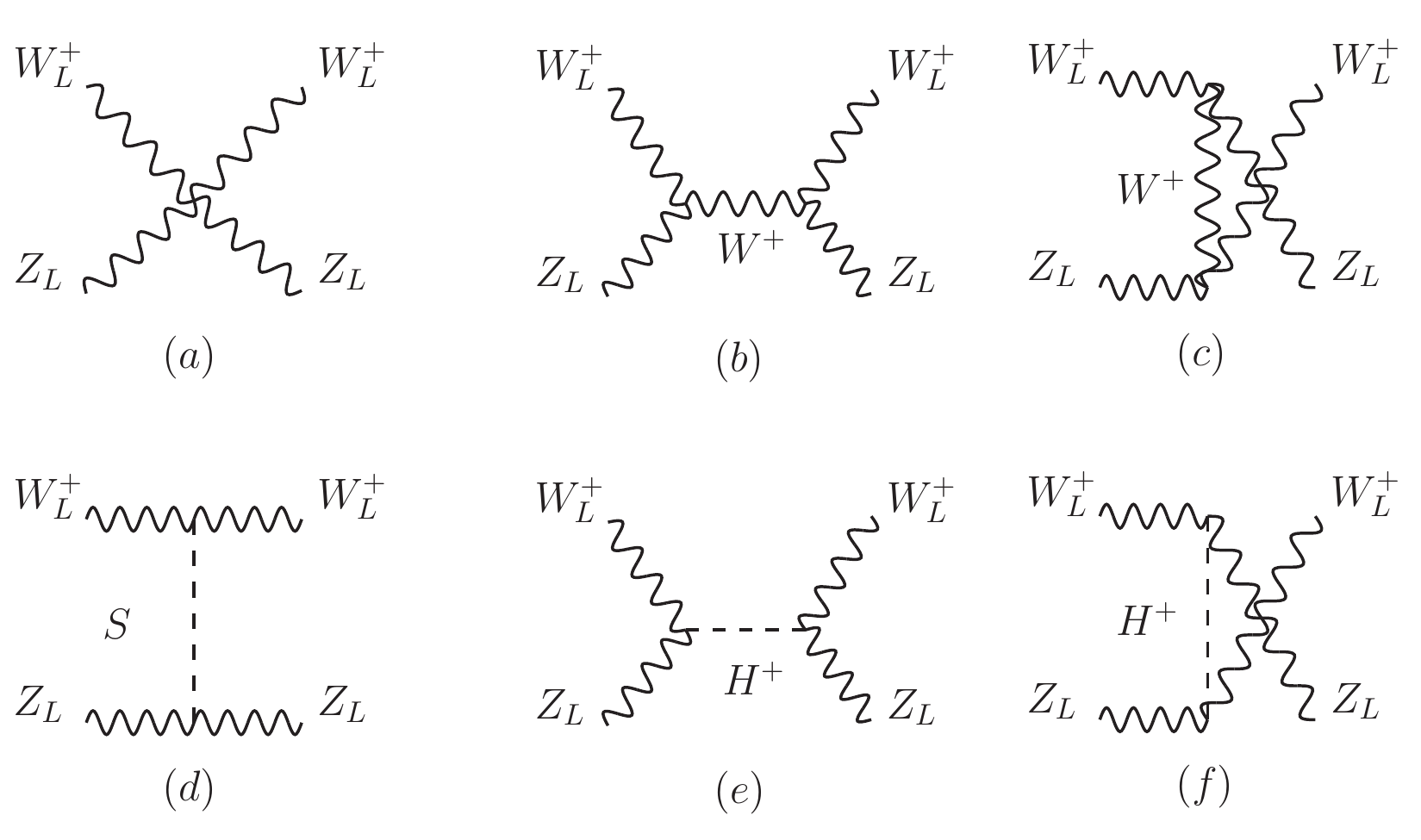}
 \caption{\label{fig:WpZWpZ} \textit{Generic Feynman diagrams for~~$W_L^+(p_1)~+~ Z_L(p_2) \rightarrow W_L^+(k_1)~+~Z_L(k_2)$~~scattering. } }
 }
 \end{center}
 \end{figure}
\bea
(a)& {\cal M}_{\rm p} &\hspace{-2ex} =~ -\frac{c_W^2 g^2}{16 E_{CM}^4 M_{W}^2 M_{Z}^2} \bigg\{4 E_{CM}^2 \left(M_{W}^2-M_{Z}^2\right)^2 \left(M_{W}^2+M_{Z}^2\right) (x-1)\nn\\
&&\hspace{2cm}+\left(M_{W}^2-M_{Z}^2\right)^4 (1-x)^2+2 E_{CM}^4 \left(M_{W}^2-M_{Z}^2\right)^2 \left(1-4 x-x^2\right)\nn\\
&&\hspace{2cm}+4 E_{CM}^6 \left(M_{W}^2+M_{Z}^2\right) (1+3 x)-E_{CM}^8 \left(3+6 x-x^2\right)\bigg\}.\nn\\
(b)& {\cal M}^{W}_{s} &\hspace{-2ex} =~  -\frac{c_W^2 g^2}{4 E_{CM}^2 M_{W}^4 M_{Z}^2(s-M_W^2)} \bigg[ E_{CM}^6 \left(M_{W}^2-M_{Z}^2\right)^2-\left(M_{W}^5-M_{W} M_{Z}^4\right)^2 (1-x)\nn\\
&&\hspace{2cm}+E_{CM}^8 M_{W}^2 x-E_{CM}^4 \big\{2 M_{Z}^6-4 M_{W}^4 M_{Z}^2 (1-2 x)-M_{W}^2 M_{Z}^4 (1-2 x)\nn\\
&&\hspace{2cm}+M_{W}^6 (3+2 x)\big\}+E_{CM}^2 \big\{M_{W}^2+M_{Z}^2)
 (3 M_{W}^6+M_{W}^2 M_{Z}^4+M_{Z}^6\nn\\
&&\hspace{2cm}-M_{W}^4 M_{Z}^2 (5+8 x)\big\}\bigg].\nn\\
(c)& {\cal M}^{W}_{u} &\hspace{-2ex} =~~ \frac{M_{Z}^2c_W^2 g^2}{32 E_{CM}^6 M_{W}^4 (u-M_W^2)}  \bigg[\big\{ M_{W}^2 (M_{W}^2-M_{Z}^2)^6 (x-1)^3+E_{CM}^{12} M_{W}^2 (-3+x) (1+x)^2\nn\\
&&\hspace{2cm}-2 E_{CM}^2 (M_{W}^2-M_{Z}^2)^4 (1-x)^2 (-M_{Z}^4+M_{W}^4 x+M_{W}^2 M_{Z}^2 (3+x)\big\}\nn\\
&&\hspace{2cm} + 2 E_{CM}^{10} \big\{M_{Z}^4 (1+x)^2+M_{W}^2 M_{Z}^2 (1+9 x+7 x^2-x^3)\nn\\
&&\hspace{2cm}+M_{W}^4 (4+15 x+10 x^2-x^3)\big\}+4 E_{CM}^6\big\{M_{Z}^8 (3-x^2)\nn\\
&&\hspace{2cm}+M_{W}^6 M_{Z}^2 (9-9 x-7 x^2-x^3)- M_{W}^8 (2-x-10 x^2-x^3)\nn\\
&&\hspace{2cm}-M_{W}^2 M_{Z}^6 (9-x-15 x^2-x^3)+M_{W}^4 M_{Z}^4 (15-9 x-17 x^2-x^3)\big\}\nn\\
&&\hspace{2cm}-E_{CM}^4 (M_{W}^2-M_{Z}^2)^2 \big\{8 M_{Z}^6 (1-x)-M_{W}^6 (7+5 x-11 x^2-x^3)\nn\\
&&\hspace{2cm}-M_{W}^2 M_{Z}^4 (23-11 x-11 x^2-x^3)+2 M_{W}^4 M_{Z}^2 (3+25 x-27 x^2-x^3)\big\}\nn\\
&&\hspace{2cm}-E_{CM}^8 \big\{8 M_{Z}^6 (1+x)+2 M_{W}^4 M_{Z}^2 (9+25 x+31 x^2-x^3)\nn\\
&&\hspace{2cm}-M_{W}^2 M_{Z}^4 (13-19 x-49 x^2-x^3)+M_{W}^6 (3+35 x+49 x^2+x^3)\big\}\bigg].\nn\\
(d)& {\cal M}^{S}_{t} &\hspace{-2ex} =~ -\frac{(C~g_2 M_W) (C^{\prime}~\frac{g_2 M_Z}{c_W})}{t-M_S^2} ~\bigg(\frac{1}{16 E_{CM}^4 M_W^2 M_Z^2}\bigg)~\bigg[\big\{(x-1)\big(E_{CM}^4+(M_W^2-M_Z^2)^2\big)\nn\\
&&\hspace{2cm}+2 E_{CM}^2 \big(M_Z^2 (1-x)+M_W^2 (1+x)\big)\big\} \big\{(x-1)\big(E_{CM}^4+(M_W^2-M_Z^2)^2\big)\nn\\
&&\hspace{2cm}+2 E_{CM}^2 \big(M_W^2 (1-x)-M_Z^2 (1+x)\big)\big\}\bigg].\nn\\
(e)& {\cal M}^{H^+}_{u} &\hspace{-2ex} =~ -\frac{(\widetilde{C}~\frac{g_2 M_Z}{c_W})^2}{u-M_{H^+}^2} ~\bigg(\frac{1}{16 E_{CM}^4 M_W^2 M_Z^2}\bigg) \bigg\{ 2 E_{CM}^2 (M_W^2+M_Z^2)-(M_W^2-M_Z^2)^2 (1-x)\nn\\
&&\hspace{10cm}-E_{CM}^4 (1+x)\bigg\}^2.\nn\\
(f)& {\cal M}^{H^+}_{s} &\hspace{-2ex} =~~  -\frac{(\widetilde{C}~\frac{g_2 M_Z}{c_W})^2}{s-M_{H^+}^2} ~\frac{(-E_{CM}^2+M_W^2+M_Z^2)^2}{4 M_W^2 M_Z^2}.\nn
\eea
\subsection{$Z_L(p_1)~+~ Z_L(p_2) \rightarrow Z_L(k_1)~+~Z_L(k_2)$}
 \begin{figure}[h!]
 \begin{center}
 {
 \includegraphics[width=16cm,height=5cm, angle=0]{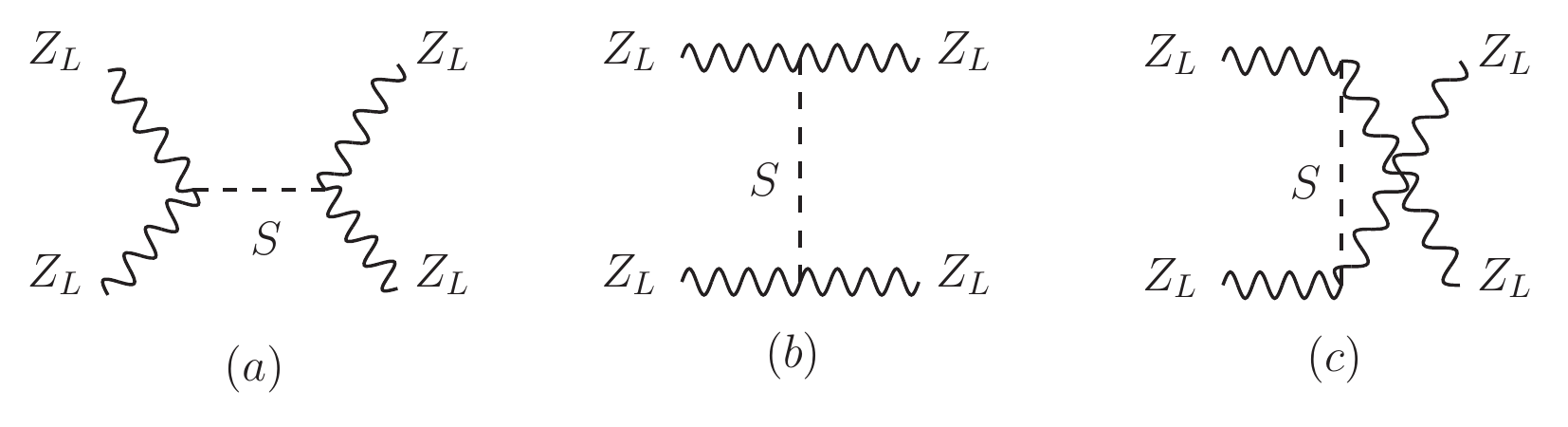}
 \caption{\label{fig:ZZZZ} \textit{Generic Feynman diagrams  for~~$Z_L(p_1)~+~ Z_L(p_2) \rightarrow Z_L(k_1)~+~Z_L(k_2)$~~scattering. } }
 }
 \end{center}
 \end{figure}
\bea
(a)& {\cal M}^{S}_{s} &\hspace{-2ex} =~~ -\frac{(C^{\prime}~\frac{g_2 M_Z}{c_W})^2}{(s-M_S^2)}~\frac{(E_{CM}^2-2 M_Z^2)^2}{4 M_Z^4}.\nn\\
(b)& {\cal M}^{S}_{t} &\hspace{-2ex} =~~ -\frac{(C^{\prime}~\frac{g_2 M_Z}{c_W})^2}{(t-M_S^2)}~\frac{\big\{4 M_Z^2+E_{CM}^2 (-1+x)\big\}^2}{16 M_Z^4}.\nn\\
(c)& {\cal M}^{S}_{u} &\hspace{-2ex} =~~ -\frac{(C^{\prime}~\frac{g_2 M_Z}{c_W})^2}{(u-M_S^2)}~\frac{\big\{-4 M_Z^2+E_{CM}^2 (1+x)\big\}^2}{16 M_Z^4}.\nn
\eea

\newpage
\setcounter{equation}{0}
\section{Total amplitude for the different modes of $V_L V_L$ scattering}\label{highenergy}

Generally when $E_{CM}\gg M_i$\,($i\equiv W,Z,h,H,H^\pm,H^{\pm\pm}$), one can express $V_L V_L\rightarrow V_L V_L$ scattering amplitude as,
\bea\label{energyseries}
{\cal M}&=& A_4~E_{CM}^4 + A_2~E_{CM}^2 + A_{0} + A_{-2}~E_{CM}^{-2}+....
\eea
The coefficient of $E_{CM}^4$ term i.e., $A_{4}$ is always zero from gauge symmetry. $A_2$ also vanishes after scalar sector contributions are added to the gauge contributions, thus ensuring unitarity of the scattering matrix. The energy independent part $A_0$ becomes the dominant one at high energy. 
It is given by 
\bea
A_{0}&=& A_{0,V} + A_{0,S}\, ,
\eea
where $A_{0,V}$ and $A_{0,S}$ are contributions from the gauge bosons and the scalars respectively. To enable the reader to readily evaluate extended scalar sector contributions to VBS processes we will now present expressions for $A_{0,V}$ and $A_{0,S}$ in various vector bosons scattering modes.

\begin{itemize}

\item $W_L^+ W_L^- \rightarrow W_L^+ W_L^-$
\bea
A_{0,V} &=& -\frac{g_2^2}{4 M_W^4 (1-x)} \bigg\{-c_W^2 M_Z^4 (3+x^2)+2 c_W^2 M_W^2 M_Z^2 (3+6 x-x^2)\nn\\&&\hspace{7.5cm} -4 M_W^4 (1+4 x-x^2)\bigg\}\nn,\\
A_{0,S} &=&-\frac{1}{4 M_W^4}\bigg\{ \big({g_2 M_W}\hat{C}\big)^2 \big(M_{H^{++}}^2+2 M_W^2 (1-x)\big) \nn\\&&\hspace{3.8cm}+2 \sum_{S=h,H} \big(g_2 {M_W} C\big)^2 \big({M_S}^2-M_W^2 (1-x)\big) \bigg \} \nn.
\eea

\item  $W_L^+ W_L^+ \rightarrow W_L^+ W_L^+$
\bea
A_{0,V} &=& - \frac{{g_2}^2}{2 M_W^4 (1-x^2)} \bigg \{c_W^2 M_Z^4 (3+x^2)+4 M_W^4 (1+3 x^2)-2 c_W^2 M_W^2 M_Z^2 (3+5 x^2)\bigg \},\nn\\
A_{0,S} &=&-\frac{1}{4 M_W^4} \bigg \{({g_2 M_W}\hat{C})^2 (M_{H^{++}}^2-4 M_W^2)+2 \sum_{S=h,H} (g_2 {M_W} C)^2 (M_S^2+2 M_W^2)\bigg\}.\nn
\eea

\item $W_L^+ W_L^- \rightarrow Z_L Z_L$
\bea
A_{0,V} &=& \frac{c_W^2 g_2^2 M_Z^2 }{2 M_W^4 (1-x^2)} \bigg \{-M_Z^2 (1-x^2)+2 M_W^2 (1+x^2)\bigg\},\nn\\
A_{0,S} &=&\frac{1}{4 M_W^2 M_Z^2}\bigg \{-2 \bigg(\frac{g_2 M_Z}{c_W}\widetilde{C}\bigg)^2 \big(M_{H^+}^2+M_W^2+M_Z^2\big)\nn\\&&\hspace{2cm} +\sum_{S=h,H} \bigg(\frac{g_2 M_Z}{c_W} C'\bigg) \big(g_2 M_W C \big) \big(-{M_S}^2+2 (M_W^2+M_Z^2)\big)\bigg \}.\nn
\eea

\item $W_L^+ Z_L \rightarrow W_L^+ Z_L$
\bea
A_{0,V} &=& \frac{c_W^2 {g_2}^2 M_Z^2 }{4 M_W^4 (1+x)}\bigg \{-2 M_W^2 (1-x)+M_Z^2 (1+x)^2 \bigg \},\nn\\
A_{0,S} &=&-\frac{1}{4 M_W^2 M_Z^2} \bigg \{\bigg(\frac{g_2 M_Z}{c_W}\widetilde{C}\bigg)^2 \big(2 M_{H^+}^2-(M_W^2+M_Z^2) (1+x)\big)\nn\\&&\hspace{2cm}+\sum_{S=h,H} \bigg(\frac{g_2 M_Z}{c_W} C'\bigg) \big(g_2 M_W C\big) \big(M_S^2+(M_W^2+M_Z^2) (1+x)\big) \bigg \}.\nn
\eea

\item  $Z_L Z_L \rightarrow Z_L Z_L$
\bea
A_{0,V} &=& 0,\nn\\
A_{0,S} &=&-\frac{3}{4M_Z^4} \sum_{S=h,H} \bigg(\frac{ g_2 M_Z}{c_W} C'\bigg)^2 M_S^2.\nn
\eea

\end{itemize}

\newpage

\setcounter{equation}{0}
\section{Required Feynman rules for vector boson scattering}\label{fyenrules}

The Feynman rules for the different vertices with the assumption that all momenta and fields are incoming.

\begin{minipage}{60mm}
 {\includegraphics[]{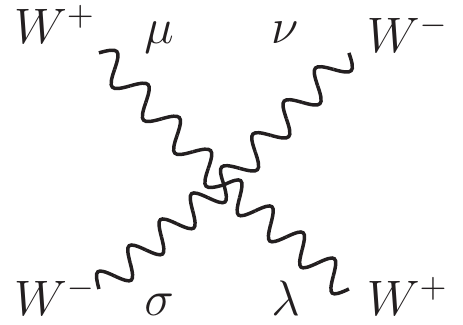}}
\end{minipage}
\begin{minipage}{10cm}
 $\displaystyle  : i g_2^2 (2 g^{\mu\lambda} g^{\sigma\nu}-g^{\mu\sigma} g^{\nu\lambda}-g^{\sigma\lambda}g^{\mu\nu}).$
\end{minipage}
\vspace{-12.8ex}\begin{equation}\end{equation}\vspace{4ex}

\begin{minipage}{60mm}
  {\includegraphics[]{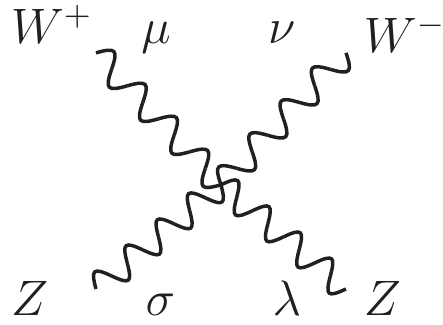}}
\end{minipage}
\begin{minipage}{10cm}
 $\displaystyle  : - i g_2^2 c_W^2(2 g^{\mu\nu} g^{\sigma\lambda}-g^{\nu\sigma} g^{\mu\lambda}-g^{\nu\lambda}g^{\mu\sigma}).$
\end{minipage}
\vspace{-12.8ex}\begin{equation}\end{equation}\vspace{4ex}

\begin{minipage}{60mm}
  {\includegraphics[]{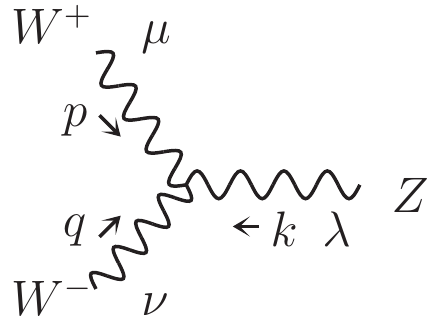}}
\end{minipage}
\begin{minipage}{10cm}
  $\displaystyle  : i g_2c_W \lbrace(p-q)^\lambda~g^{\mu\nu}+(q-k)^\mu~g^{\lambda\nu}+(k-p)^\nu~g^{\mu\lambda}\rbrace.$
\end{minipage}
\vspace{-12.8ex}\begin{equation}\end{equation}\vspace{4ex}

\begin{minipage}{60mm}
 {\includegraphics[]{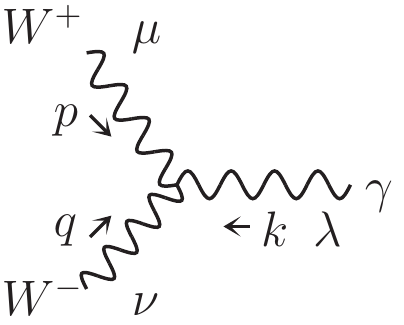}}
\end{minipage}
\begin{minipage}{10cm}
  $\displaystyle  : i g_2s_W \lbrace(p-q)^\lambda~g^{\mu\nu}+(q-k)^\mu~g^{\lambda\nu}+(k-p)^\nu~g^{\mu\lambda}\rbrace.$
\end{minipage}
\vspace{-12.8ex}\begin{equation}\end{equation}\vspace{4ex}

\begin{minipage}{60mm}
  {\includegraphics[]{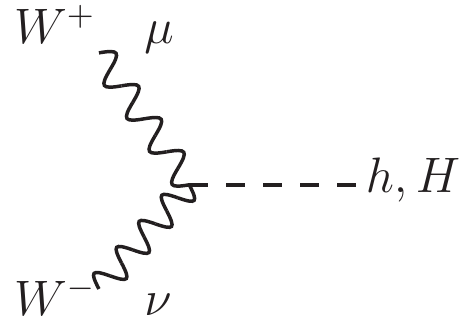}}
\end{minipage}
\begin{minipage}{10cm}
  $\displaystyle  : {i g_2}{M_W} g_{\mu\nu} C$, where $C$ is given by:
\end{minipage}

\vspace{-6ex}
\begin{equation}
\begin{aligned}
  &  {\rm SM}  &
  &:\left\{\begin{array}{l}{\rm for}~~h,~~C = 1,\\
      {\rm for}~~H,~~C = 0,\end{array}\right.\\
  & {\rm 2HDM}  &
  &:\left\{\begin{array}{l}{\rm for}~~h,~~C = \sin(\beta-\alpha),\\
      {\rm for}~~H,~~C = \cos(\beta-\alpha),\end{array}\right.\\
  & {\rm Y=0~~HTM} &
  &:\left\{\begin{array}{l}{\rm for}~~h,~~C = (\cos\widetilde{\beta}\cos{\widetilde{\gamma}}+2\sin\widetilde{\beta}\sin{\widetilde{\gamma}}),\\
      {\rm for}~~H,~~C = (-\cos\widetilde{\beta}\sin{\widetilde{\gamma}}+2\sin\widetilde{\beta}\cos{\widetilde{\gamma}}),\end{array}\right.\\
  & {\rm Y=2~~HTM} &
  &:\left\{\begin{array}{l}{\rm for}~~h,~~C = (\cos{\beta'}\cos{\gamma'}+\sqrt{2}\sin{\beta'}\sin{\gamma'}),\\
      {\rm for}~~H,~~C = (-\cos{\beta'}\sin{\gamma'}+\sqrt{2}\sin{\beta'}\cos{\gamma'}).\end{array}\right.
\end{aligned}
\end{equation}

\begin{minipage}{60mm}
  {\includegraphics[]{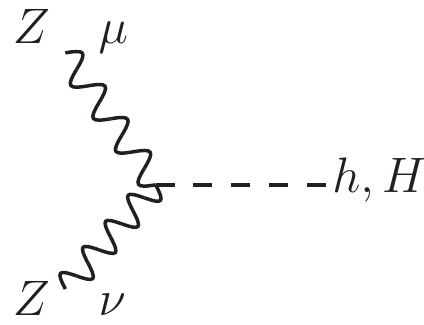}}
\end{minipage}
\begin{minipage}{10cm}
  $\displaystyle  : \frac{i g_2 M_Z}{c_W} g_{\mu\nu} C'$, where $C'$ is given by:
\end{minipage}

\vspace{-6ex}
\begin{equation}
\begin{aligned}
  &  {\rm SM} &:
  &\left\{\begin{array}{l}{\rm for}~~h,~~C' = 1,\\
      {\rm for}~~H,~~C' = 0,\end{array}\right.\\
  & {\rm 2HDM}  &:
  &\left\{\begin{array}{l}{\rm for}~~h,~~C' = \sin(\beta-\alpha),\\
      {\rm for}~~H,~~C' = \cos(\beta-\alpha),\end{array}\right.\\
  & {\rm Y=0~~HTM}  &:
  &\left\{\begin{array}{l}{\rm for}~~h,~~C' = \cos\widetilde{\gamma},\\
      {\rm for}~~H,~~C' = -\sin{\widetilde{\gamma}},\end{array}\right.\\
  & {\rm Y=2~~HTM}  &:
  &\left\{\begin{array}{l}{\rm for}~~h,~~C' = (\cos{\delta'}\cos{\gamma'}+2\sin{\delta'}\sin{\gamma'}),\\
      {\rm for}~~H,~~C' = (-\cos{\delta'}\sin{\gamma'}+2\sin{\delta'}\cos{\gamma'}).\end{array}\right.
\end{aligned}
\end{equation}


\begin{minipage}{60mm}
  {\includegraphics[]{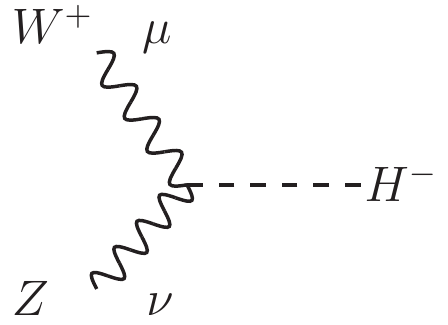}}
\end{minipage}
\begin{minipage}{10cm}\vspace{3ex}
$\displaystyle ~~~: \frac{i g_2 M_Z}{c_W} g_{\mu\nu} \widetilde{C}$, where $\widetilde{C}$ is given by:
\end{minipage}

\allowdisplaybreaks
\begin{equation}
 \begin{aligned}
  ~~~~~~~~~~~~~~~&{\rm SM}:  \hspace{3cm} \widetilde{C}   =0,\\ 
  &{\rm 2HDM}: \hspace{2.4cm} \widetilde{C}   =0,\\
  &Y=0{\rm ~~HTM}: \hspace{1.3cm} \widetilde{C}   = \sin\widetilde{\beta}\cos\widetilde{\beta}~\frac{M_W}{M_Z},\\
  &Y=2{\rm ~~HTM}: \hspace{1.3cm} \widetilde{C}   = c_W\bigg\lbrace\sin{\beta'}\cos{\delta'}s_W^2-\frac{(1+s_W^2)}{\sqrt{2}}\cos{\beta'}\sin{\delta'} \bigg\rbrace.  
 \end{aligned}
\end{equation}

\begin{minipage}{60mm}
  {\includegraphics[]{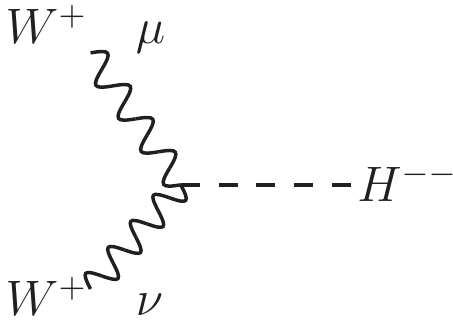}}
\end{minipage}
\begin{minipage}{10cm}\vspace{3ex}
$\displaystyle ~~~: {i g_2 M_W}g_{\mu\nu} \hat{C}$, where $\hat{C}$ is given by:
\end{minipage}

\allowdisplaybreaks
\begin{equation}
 \begin{aligned}
 \hspace{-2cm} ~&{\rm SM}: \hspace{3cm} \hat{C}   =0,\\ 
  &{\rm 2HDM}: \hspace{2.4cm} \hat{C}   =0,\\
  &Y=0{\rm ~~HTM}: \hspace{1.3cm}\hat{C}   = 0,\\
  &Y=2{\rm ~~HTM}: \hspace{1.3cm} \hat{C}   = 2\sin{\beta'}.\qquad \qquad\qquad\qquad\qquad\qquad ~~~~~  
 \end{aligned}
\end{equation}

\end{appendices}

\end{document}